\documentclass[11pt,oneside,reqno,reprint,twocolumn,nofootinbib]{revtex4-1}
\usepackage{amsfonts}
\usepackage[a4paper,left=1in,right=1in,top=1in,bottom=1in,footskip=.25in]{geometry}
\usepackage{graphicx}
\usepackage{subcaption}
\usepackage[noabbrev]{cleveref}
\usepackage{amsmath}
\usepackage{amssymb}
\usepackage{epstopdf}
\usepackage{color}

\begin{document}

\title{Cyclic Mixmaster Universes}
\author{John D. Barrow}
\email{J.D.Barrow@damtp.cam.ac.uk}
\author{Chandrima Ganguly}
\email{C.Ganguly@damtp.cam.ac.uk}
\affiliation{DAMTP, Centre for Mathematical Sciences,\\
University of Cambridge,\\
Wilberforce Rd.,\\
Cambridge CB3 0WA\\
United Kingdom}
\date{17th March, 2017}

\begin{abstract}
We investigate the behaviour of bouncing Bianchi type IX `Mixmaster'
universes in general relativity. This generalises all previous studies of
the cyclic behaviour of closed spatially homogeneous universes with and
without entropy increase. We determine the behaviour of models containing
radiation by analytic and numerical integration and show that increase of
radiation entropy leads to increasing cycle size and duration. We introduce
a null energy condition violating ghost field to create a smooth,
non-singular bounce of finite size at the end of each cycle and compute the
evolution through many cycles with and without entropy increase injected at
the start of each cycle. In the presence of increasing entropy we find that
the cycles grow larger and longer and the dynamics approach flatness, as in
the isotropic case. However, successive cycles become increasingly
anisotropic at the expansion maxima which is dominated by the
general-relativistic effects of anisotropic 3-curvature. When the dynamics
are significantly anisotropic the 3-curvature is negative. However, it
becomes positive after continued expansion drives the dynamics close enough
to isotropy for the curvature to become positive and for gravitational
collapse to ensue. In the presence of a positive cosmological constant,
radiation and a ghost field we show that, for a very wide range of
cosmological constant values, the growing oscillations always cease and the
dynamics subsequently approach those of the isotropic de Sitter universe at
late times. This model is not included in the scope of earlier cosmic
no-hair theorems because the 3-curvature can be positive. In the case of
negative cosmological constant, radiation and an ultra-stiff field (to
create non-singular bounces) we show that a sequence of chaotic oscillations
also occurs, with sensitive dependence on initial conditions. In all cases,
we follow the oscillatory evolution of the scale factors, the shear, and the
3-curvature from cycle to cycle.
\end{abstract}

\maketitle

\section{ Introduction}

In 1922 Alexander Friedmann \cite{fried} first noted the existence of
`periodic worlds' in his solutions of Einstein's equations for isotropic and
homogeneous universes with positive spatial curvature. But the physical
study of cyclic universes in general relativistic cosmology begins with the
work of Tolman \cite{tol}, who first considered the simple situation of a
closed Friedmann universe with zero cosmological constant, $\Lambda $, and
non-negative pressure. The evolution of Tolman's cyclic universes can be
continued periodically through the big crunch singularity if it is assumed
that no new physics arises there and the evolution can be extended smoothly
through it \footnote{%
Tolman's analysis, which assumed no equation of state linking the pressure, $%
p\geq 0$, to the density, $\rho $, required an additional assumption. To
avoid a finite-time (`sudden') singularity occuring where $p\rightarrow
\infty $ with finite $\rho $ before the expansion maximum is reached one
must stipulate some control over $p,$ for example $p<C\rho $ for constant $%
C>0$ \cite{bgt}.}. As a next step, Tolman incorporated the general
consequences of an increase of entropy from cycle to cycle, in accordance
with the second law of thermodynamics. This produced a monotonic increase in
the maximum size and length of successive cycles, which continues forever.
If this entropy increase is modelled as an increase in the dimensionless
entropy per baryon in a mixture of radiation and baryons (ignoring baryon
non-conserving interactions) then the increase is a reflection of the
asymmetry in the pressure, $p$, from cycle to cycle, as energy is
transferred from the baryonic ($p=0$) to the radiation ($p\neq 0$) gas.
Tolman's work attracted periodic interest by other astronomers like Bonnor 
\cite{bonn} and Zanstra \cite{zan} in the 1950s, before Zeldovich and
Novikov turned Tolman's result into a theorem for rather restrictive
equations of state of matter obeying Friedmann's equations and the laws of
thermodynamics. They also assumed that the cycles could not be continued
indefinitely into the past because they would become smaller than the
smallest finite-sized elementary particles (not assumed pointlike in those
days). As the cycles continue to increase in size, an oscillating universe
appears increasingly `flat', although it is closed with positive spatial
curvature \cite{LPark}. This might even provide an explanation for the
proximity of the expansion dynamics to flatness today that differs in detail
from that of the standard one-cycle inflationary universe model (although
the latter generates proximity to flatness by a large entropy increase in
one cycle through `reheating' rather than by a progressive build up over
many cycles by all processes). However, in what follows we will show that it
does not share other features of an inflationary universe at late times.

Generalisations of this simple oscillatory Tolman universe produced some
interesting new features. Barrow and Dabrowski \cite{bdab} show that if
there is a positive cosmological constant ($\Lambda >0$) then the sequence
of growing oscillations always comes to an end after a finite proper time
and the dynamics evolve towards an ever-expanding de Sitter asymptote as $%
t\rightarrow \infty $ \cite{bdab}. This end to the oscillations occurs no
matter how small the positive value of the cosmological constant is: the
cycles grow until they inevitably produce one that is large enough for the $%
\Lambda $ term to eventually dominate the dynamics at large size and its
effect is to stop a further contraction from occurring. Notice that the
final state is always one which is close to flatness and only just
(depending on the exact size of the inter-cycle entropy jump) dominated by
the $\Lambda $ energy density -- rather like our universe, in fact.

Our paper is organised as follows. We begin in section \ref%
{sec:isotropic_bounces} by presenting some simple exact solutions with
finite minima that illustrate the usefulness of introducing ghost fields
into any model of bouncing universes. In section \ref{sec:the_model} we
describe the thermodynamic aspects of the evolution. In section \ref%
{sec:numerics} we give diagonal Bianchi type IX metric and field equations
and present some simple approximate parametric solutions for the case of the
radiation-dominated axisymmetric type IX universe, as well as a solution for
the purely cosmological constant dominated axisymmetric type IX universe (a
case not encompassed by the cosmic no-hair theorems). These will serve to
guide our interpretation of the numerical solutions of the full dynamical
equations. In the next section we present the results of numerically solving
the full Bianchi type IX equations, first during a single cycle without the
ghost field, then with the ghost field over many cycles. We study cases with
and without the injection of radiation entropy. Finally, we study the same
problems with the addition of positive or negative cosmological constants.

\section{\label{sec:isotropic_bounces}Simple isotropic bounces}

The assumption of a `bounce' occurring at zero expansion scale and infinite
density is computationally (and physically) awkward. However it can be
improved upon by introducing a simple `ghost' field, with negative density, $%
\rho <0$ , that will cause the expansion to go through a smooth minimum at
the beginning and at the end of each cycle instead of through a singularity
where $\rho =\infty $. Ghost fields have often been used in bouncing
cosmology scenarios to effect a non singular bounce such as in \cite%
{cai1,cai2} As illustrations, we can find two simple exact solutions which
are of use in more complicated situations. Suppose that we have a closed
Friedmann universe with scale factor, $a(t)$, containing two `fluids' having
densities $\rho >0$ and $\rho _{g}<0$. The second `ghost' fluid with
negative density, $\rho _{g}$, acts as a model stress to dominate at small $%
a $ and effect a bounce at $a=a_{\min }$, while the conventional fluid with
positive density, $\rho $, dominates at larger $a$. This situation continues
to exist until the spatial curvature creates an expansion maximum at $%
a=a_{\max }.$ We give two solutions which are useful models of this type of
behaviour for more detailed analyses and illustrate the effects of the two
fields:

\textit{Ghost fluid with }$p_{g}=\rho _{g}\propto a^{-6}<0$\textit{\ and
conventional radiation fluid with }$p=\rho /3\propto a^{-4}>0$

The Friedmann equation (setting $8\pi G=c=1$) is

\begin{equation*}
\frac{\dot{a}^{2}}{a^{2}}=-\frac{\Sigma}{a^{6}}+\frac{\Gamma }{a^{4}}-\frac{1%
}{a^{2}},
\end{equation*}%
with $\Sigma \geq 0$ and $\Gamma \geq 0$ being constants. The exact solution
for the scale factor, when $\Gamma ^{2}\geq 4\Sigma $, can be written simply
in terms of the expansion maximum and minimum radii in conformal time,
defined by $dt=ad\eta $, as \cite{Btsag}

\begin{equation*}
a^{2}(\eta )=\frac{1}{2}\left[ a_{\max }^{2}+a_{\min }^{2}+(a_{\max
}^{2}-a_{\min }^{2})\sin 2(\eta +\eta _{0})\right] ,\ 
\end{equation*}%
where the integration constant $\eta _{0}$ can be set to zero without loss
of generality and

\begin{eqnarray*}
a_{\min }^{2} &\equiv &\frac{\Gamma -\sqrt{\Gamma ^{2}-4\Sigma }}{2}, \\
a_{\max }^{2} &\equiv &\frac{\Gamma +\sqrt{\Gamma ^{2}-4\Sigma }}{2}.
\end{eqnarray*}%
Oscillatory solutions occur when $\Gamma ^{2}>4\Sigma $ and $\Gamma
^{2}=4\Sigma $ gives a static universe. In the high radiation entropy ($%
\propto \rho ^{3/4}\propto \Gamma ^{3/4}$) limit $\Gamma ^{2}\gg4\Sigma ,$
we have $a_{\max }\rightarrow \Gamma $ and $a_{\min }\rightarrow \Sigma
/\Gamma $ and we see that the maxima grow and the minima decrease in size if
we let the radiation entropy grow from cycle to cycle.

Ghost fluid with $p_{g}=\rho _{g}/3\propto a^{-4}<0$ and conventional dust
fluid with $p=0,\rho \propto a^{-3}>0$

The Friedmann equation is

\begin{equation*}
\frac{\dot{a}^{2}}{a^{2}}=-\frac{\Gamma }{a^{4}}+\frac{M}{a^{3}}-\frac{1}{%
a^{2}},
\end{equation*}%
and, if $M^{2}\geq 4\Gamma ,$ a new exact solution can be written simply in
conformal time $dt=ad\eta $ in terms of the expansion maximum and minimum
radii as \cite{Btsag}

\begin{equation*}
a(\eta )=\frac{1}{4}\left[ a_{\max }+a_{\min }+(a_{\max }-a_{\min })\sin
(\eta +\eta _{0})\right] \ ,
\end{equation*}%
where

\begin{equation*}
a_{\min }=\frac{1}{2}\left[ M-\sqrt{M^{2}-4\Gamma }\right] ,\ 
\end{equation*}

\begin{equation*}
a_{\max }=\frac{1}{2}\left[ M+\sqrt{M^{2}-4\Gamma }\right] .
\end{equation*}%
In the limit that $M^{2}\gg4\Gamma $, we have $a_{\min }\rightarrow \Gamma
/M $ and $a_{\max }\rightarrow M$, so if we introduce an increase in matter
entropy ($\propto M$) from cycle to cycle then we will have successively
increasing maxima and decreasing minima.

In what follows, we shall add a ghost field to an oscillating anisotropic,
spatially homogeneous universe in order to produce a smooth bounce at finite
values of the scale factor where the densities are non-singular. It would be
possible to effect a smooth bounce with a scalar field with quadratic
potential which has been investigated in Mixmaster universes \cite{bmatz1},
however the probability of this bounce occurring is small, $O(a_{\min
}/a_{\max })$, in any universe with $a_{\min }\ll a_{\max }$.

All of the above discussion has focussed upon simple isotropic closed
universes with $S^{3}$ spatial topology. The situation in simple anisotropic
universes of Kantowski-Sachs type was studied in detail by Barrow and D\c{a}%
browski \cite{bdab} and produces a more complicated scenario for cycle to
cycle evolution. The Kantowski-Sachs universes have special $S^{2}\times
S^{1}$ spatial topology and are far from generic even amongst homogeneous
anisotropic universes, although they have inhomogeneous generalisations with
no symmetries found by Szekeres \cite{sz}. Only closed (compact space
sections) universes with $S^{2}\times S^{1}$ or $S^{3}$ topologies possess
maximal hypersurfaces and so can recollapse and bounce when gravitationally
attractive matter is present \cite{bt}. Whether or not they will do so
depends on the matter content of the universe.

In this paper we will study the dynamics of a cyclic Bianchi type IX
`Mixmaster' universe with $S^{3}$ spatial topology. This is the most general
spatially homogeneous anisotropic closed universe and it contains the closed
Friedmann universe as an isotropic special case. Exact solutions are only
known for the axisymmetric special cases in vacuum, containing stiff matter (%
$p=\rho $) or electromagnetic fields, or a combination of both. \cite{exact}%
. We are particularly interested in the behaviour of these anisotropic
universes on approach to an expansion maximum of the volume and the
behaviour of the three expansion scale factors there. Do the cycles grow in
maximum size and do they become increasingly anisotropic from cycle to
cycle? We will confine our attention to the case where the fluid in the
universe is comoving, although in a subsequent study we will generalise this
to fluids with non-comoving velocities. The type IX universe behaves quite
differently to the simple Bianchi type I anisotropic universe because it has
both expansion anisotropy (shear) and 3-curvature anisotropy. The
3-curvature anisotropy has no Newtonian analogue. The 3-curvature dynamics
are complicated and the sign of the 3-curvature varies in time and is only
positive when the expansion dynamics are sufficiently close to isotropy. An
expansion maximum will only occur when the 3-curvature, $^{(3)}R$, becomes
positive (as it is all the time in the closed Friedmann universes). We want
to discover if increasing entropy increases the size of successive expansion
maxima, as in the cyclic Friedmann models, but also determine what happens
to the expansion anisotropy over successive cycles. We can also incorporate
a positive or negative cosmological constant to see if it can terminate a
sequence of oscillations in a cyclic type IX universe in the same way that
it does in an isotropic closed universe.

In order to follow the Bianchi type IX evolution smoothly from cycle to
cycle, we introduce a stiff ghost field to create an expansion minimum at
non-zero volume in every cycle, as discussed above. This field has no
significant effect on the expansion maxima or the behaviour of the dynamics
in its vicinity. It is well known that the Bianchi IX model displays formal
chaotic behaviour in all its degrees of freedom as the volume tends to zero 
\cite{jbchaos}. However, there is only an unbounded number of chaotic
oscillations of the scale factor on an open interval $0<t<T $ around the
time origin for finite $T$: an infinite number of scale factor oscillations
occur on any such interval no matter how small the value of $T$. In any
finite interval $T_{1}<t<T,$ not including $t=0$, the number of oscillations
is finite and not technically chaotic. For realistic choices of $%
T_{1}\approx 10^{-43}s$ for the start of classical cosmology, there will be
less than about $12$ Mixmaster oscillations even if they continued all the
way from $T_{1}$ up to the present day \cite{ZN}. This is because the
overall expansion scale changes rapidly with the number of scale factor
oscillations, which occur in logarithmic time. Thus, if there is a bounce at 
$finite$ volume the issue of chaotic Mixmaster oscillations \cite{bkl} is
irrelevant to a discussion of the long-term dynamics.

Some bouncing cosmologies deal with the situation at bounce by incorporating
a so-called phase of ekpyrosis where there is effectively an ultra stiff
isotropic fluid with a $p\gg \rho $ field added to the matter content of the
universe \cite{ekpy}. If there are no anisotropic pressures, this drives the
dynamics towards isotropy as a singularity is approached. However, we should
expect anisotropic pressures to be larger than the energy density (as is
assumed for the isotropic pressure), due to the dominance of collisionless
particles when $T>10^{15}GeV$. The presence of these anisotropic pressures
during the phase of ekpyrosis can reinstate the distorting effects of
anisotropies and they diverge on approach to the bounce \cite{by,bg}.

In this work, we shall consider the effect of expansion and curvature
anisotropies on a model of a bouncing type IX universe. This model
incorporates several bounces, and we increase the entropy of the constituent
matter content via an injection at each bounce. The increased entropy in
each bounce leads to a higher maxima and longer cycles, as found in the
original analysis of Tolman. The claim is that with increasing maxima,
simple isotropic bouncing models, such as the isotropic Friedmann universe
approaches flatness and begins to resemble the present-day universe in this
respect. We shall investigate this claim in more general circumstances. We
confirm the increasing volume maxima in successive bounces and the eventual
cessation of bounces in the presence of a cosmological constant but, with a
more complicated transitional evolution for the isotropisation of the three
scale factors. However, we find that in the absence of a cosmological
constant successive cycles become increasingly anisotropic despite the
increase in size and approach to flatness. This is quite different to the
long-term evolution predicted by inflation.

We also investigate the effects of a negative cosmological constant. This
always produces collapse to a future singularity \cite{fjt}. Our aim in
doing this is to construct an anisotropic version of the simple Friedmann
universes which can all be transformed into simple harmonic oscillators in
conformal time after rescaling the expansion scale factor \cite{jbSHO}.
These models are studied as a further simple example of a bouncing type IX
universe. They include `domain-wall' matter ($p=-2/3\rho $), as well as a
negative cosmological constant in a closed Friedmann universe. We can also
include an ultra-stiff matter field with isotropic pressures ($p=5\rho $) to
subdue the anisotropies on approach to the bounce. In the absence of
pressure anisotropies, this should work and allow our model to propagate
further without hitting a singularity. In a later study we will include both
non-comoving velocities and associated pressure anisotropies into the
analysis of cyclic type IX universes.

\subsection{Entropy}

Our next task is to follow the consequences of a growth of entropy in the
constituents of the system. The definition of a cosmological entropy is
still debated. Here, we shall use only the thermodynamic entropy of the
radiation or matter content and ignore any contribution from a
`gravitational entropy' that might be associated with Weyl curvature,
gravitational clustering, or the area of the particle horizon \cite{gravS}.
We will use a very simple toy model of entropy injection in our model of a
bouncing universe. We are, in this analysis, not interested in the physical
origin of the production of entropy by non-equilibrium processes like
quantum particle production or viscous anisotropy damping, which are
dramatic entropy producers as $t\rightarrow 0$, \cite{bmatz}. Rather, we
will consider sudden entropy increase at each expansion minimum and also use
the dependence of the size of the expansion maximum on the entropy to
determine the effect of increasing the entropy in a cycle. To circumvent
issues regarding the non-conservation of baryon number, and hence making the
definition of entropy per baryon ambiguous, we shall consider the effects of
an increase of entropy of radiation. To model this, we consider first the
definition of the entropy of radiation, $S$, which is given by, 
\begin{equation}
S\propto T^{3}V,
\end{equation}%
where $T$ is the temperature at that instant and $V$ is the volume of the
universe. We assume that the entropy per unit volume, once injected at the
minima, remains constant throughout the duration of the cycle until the next
minimum (we ignore all particle-antiparticle annihilations and massive
particles that become non-relativistic). The radiation energy density varies
as $\rho _{r}=C_{r}V^{-4/3}$. Hence, during each cycle, the quantity $T^{3}V$
is a constant, implying that $\rho _{r}\propto T^{4}$. So, we can write the
entropy as 
\begin{equation}
S\propto T^{3}\left( \frac{C_{r}}{\rho _{r}}\right) ^{3/4}\propto
C_{r}^{3/4}.
\end{equation}%
Thus, we can assess the effect of increasing the entropy of radiation by an
increase in the constant $C_{r}$. It has been shown previously, in works
such as \cite{tol}, that an increase in the entropy of radiation leads to an
increase in the expansion maxima in closed Friedmann universes. \emph{\ }We
expect this to occur also for a Bianchi IX universe. However, we also wish
to study how the shape of the anisotropy behaves as the cycles get bigger
with entropy injection. For this we need to specify the form of the type IX
metric and analyze the field equations.

\section{\label{sec:the_model}The model of the bouncing Bianchi type IX
universe}

\subsection{Einstein equations for diagonal type IX universes}

We aim to study the effects of entropy growth in anisotropic oscillating
models of Bianchi type IX. We divide the problem into several sub-cases. We
are interested in discovering whether the present-day universe would
isotropise and approach flatness in a bouncing universe model after many
cycles of entropy growth. In all cases, we take our cosmological \ model to
be the spatially homogeneous Bianchi Type IX spacetime containing radiation
with pressure $p=\rho /3,$ and\ a `dust' matter field with equation of state 
$p=0$. We use the radiation field to assess the effects of increasing the
entropy of radiation from cycle to cycle in a bouncing universe. If only
these two fluids are present, the universe will evolve towards a strong
curvature singularity and experience only one cycle unless we assume a
periodic continuation through the singularity. If we add the ultra-stiff
ghost field with $\rho_g<0$ and $p_{g}/ \rho _{g}\gg 1$ then we create a
smooth non-singular bounce and can follow the evolution through several
cycles. We can then introduce a growth of entropy in the radiation field in
the vicinity of the bounce in order to study the long-term effects of the
shear anisotropy and the $3$-curvature on the expansion maximum. In the last
section of the paper, we introduce a positive cosmological constant and also
a negative cosmological constant, to see how the evolution is changed by
their presence. In the latter case, we will add a domain- wall fluid ($%
p=-2\rho /3$) and an ultra-stiff fluid ($p=5\rho $) to facilitate smooth
non-singular bounces.

Note that when a ghost field is added to create a non-singular bounce it
means that the dynamics will be dominated by this isotropic matter field at
the expansion minima and so it will have a small isotropising effect over
small time intervals around the minima. However, it is outweighed by the
lengthening of the evolution time produced by the growing size of successive
maxima. It would be possible to effect non-singular bounces with an
anisotropic ghost field but this complication has been avoided here.

We study the solution of the Einstein equations for a diagonal spatially
homogeneous Bianchi type IX universe with metric \cite{LL}

\begin{eqnarray}
ds^{2} &=&dt^{2}-\gamma _{ab}(t)e_{\mu }^{a}e_{\nu }^{b}dx^{\mu }dx^{\nu },
\label{metric} \\
\gamma _{ab}(t) &=&diag[a^{2}(t),b^{2}(t),c^{2}(t)],  \notag \\
e_{\mu }^{a} &=&%
\begin{bmatrix}
\cos z & \sin z\sin x & 0 \\ 
-\sin z & \cos z\sin x & 0 \\ 
0 & \cos x & 1%
\end{bmatrix}%
,  \notag
\end{eqnarray}

containing non-interacting perfect fluids, each with a perfect fluid
equation of state $p=(\gamma -1)\rho .$ The equations of state parameters
for the radiation, dust and ghost fields are given by $\gamma _{r}=4/3$, $%
\gamma _{m}=1$ and $\gamma _{g}=2,$ respectively. The orthogonal expansion
scale factors $a(t)$, $b(t)$ and $c(t)$ for the Bianchi type IX universe
with the specified matter content satisfy the field equations: 
\begin{equation}
\begin{split}
\frac{\ddot{a}}{a}+\frac{\ddot{b}}{b}+\frac{\dot{a}\dot{b}}{ab}+\frac{a^{2}}{%
4b^{2}c^{2}}+\frac{b^{2}}{4a^{2}c^{2}}-\frac{3c^{2}}{4a^{2}b^{2}}+\frac{1}{%
2a^{2}}+\frac{1}{2b^{2}} \\
-\frac{1}{2c^{2}}=-\sum_{i=r,m,g}(\gamma _{i}-1)\rho ,  \label{11}
\end{split}%
\end{equation}
\begin{equation}
\begin{split}
\frac{\ddot{b}}{b}+\frac{\ddot{c}}{c}+\frac{\dot{b}\dot{c}}{bc}+\frac{b^{2}}{%
4a^{2}c^{2}}+\frac{c^{2}}{4a^{2}b^{2}}-\frac{3a^{2}}{4b^{2}c^{2}}+\frac{1}{%
2b^{2}}+\frac{1}{2c^{2}} \\
-\frac{1}{2a^{2}} =-\sum_{i=r,m,g}(\gamma _{i}-1)\rho ,  \label{22}
\end{split}%
\end{equation}
\begin{equation}
\begin{split}
\frac{\ddot{c}}{c}+\frac{\ddot{a}}{a}+\frac{\dot{c}\dot{a}}{ca}+\frac{a^{2}}{%
4b^{2}c^{2}}+\frac{c^{2}}{4a^{2}b^{2}}-\frac{3b^{2}}{4a^{2}c^{2}}+\frac{1}{%
2a^{2}}+\frac{1}{2c^{2}} \\
-\frac{1}{2b^{2}} =-\sum_{i=r,m,g}(\gamma _{i}-1)\rho .  \label{33}
\end{split}%
\end{equation}
The constraint equation reduces to,

\begin{equation}
\begin{split}
\frac{\dot{a}\dot{b}}{ab}+\frac{\dot{b}\dot{c}}{bc}+\frac{\dot{c}\dot{a}}{ca}%
+\frac{1}{2a^{2}}+\frac{1}{2b^{2}}+\frac{1}{2c^{2}}-\frac{a^{2}}{4b^{2}c^{2}}
\\
-\frac{b^{2}}{4a^{2}c^{2}}-\frac{c^{2}}{4a^{2}b^{2}}=\sum_{i=r,m,g}\rho
\label{00}
\end{split}%
\end{equation}

From the fluid continuity equations, we have, 
\begin{align*}
\rho _{r}(t)& \propto (abc)^{-4/3}\  \\
\ \rho _{g}(t)& \propto (abc)^{-2} \\
\ \rho _{m}(t)& \propto (abc)^{-1}
\end{align*}

If we introduce a new time coordinate, $\tau $, by defining 
\begin{equation}
d\tau =dt/abc  \label{tau}
\end{equation}

then the field equations become ($^{\prime }$ denotes $d/d\tau $): 
\begin{align}
2(\mathrm{ln}\,a)^{\prime \prime \ }+a^{4}-(b^{2}-c^{2})^{2}&
=a^{2}b^{2}c^{2}\sum_{i=r,m,g}(\rho _{i}-p_{i}),  \label{a} \\
2(\mathrm{ln}\,b)^{\prime \prime \ }+b^{4}-(c^{2}-a^{2})^{2}&
=a^{2}b^{2}c^{2}\sum_{i=r,m,g}(\rho _{i}-p_{i}),  \label{b} \\
2(\mathrm{ln}\,c)^{\prime \prime \ }+c^{4}-(a^{2}-b^{2})^{2}&
=a^{2}b^{2}c^{2}\sum_{i=r,m,g}(\rho _{i}-p_{i}),  \label{c}
\end{align}%
and the constraint equation simplifies to, 
\begin{equation}
\begin{split}
4[(\mathrm{ln}\,a)^{\prime }(\mathrm{ln}\,b^{\prime })+(\mathrm{ln}%
\,b)^{\prime }(\mathrm{ln}\,c)^{\prime }+(\mathrm{ln}\,c)^{\prime }(\mathrm{%
ln}\,a)^{\prime }] \\
=a^{4}+b^{4}+c^{4}-2c^{2}(a^{2}+b^{2}) -2a^{2}b^{2} \\
+4a^{2}b^{2}c^{2}\sum_{i=r,m,g}\rho _{i}.  \label{con}
\end{split}%
\end{equation}

Before we attempt to study the full numerical evolution of the type IX
equations of motion with an increase in entropy of the radiation field in
the presence of the ultra-stiff ghost field and a dust field, we shall try
to construct an approximate parametric solution for the type IX evolution
containing only the radiation field.\emph{\ }A detailed study of the
behaviour of the most general Bianchi type universes at intermediate times
has been conducted in refs. \cite{DLN, Luk}. What they reveal is that at a
very early time there is a reduction in anisotropy by quantum effects which
is significant. To the future of such a time, the evolution enters a long
quasi-axisymmetric phase. Two scale factors are larger than the third and
differences between the first two are insignificant compared to their size
relative to the other. This situation is familiar from the evolutionary
pattern during isotropisation of Kasner metrics containing collisionless
particles \cite{starob} where the anisotropic pressures created by the
particles mimic the 3-curvature anisotropies in type IX. It is as if the
dynamics has entered a time-reverse of one of the long cycles that a Bianchi
type IX universe encounters on approach to small times. \emph{\ }Following
Doroshkevich et al\emph{\ }\cite{DLN}\emph{, }we use the approximation $%
a=b\gg c,$ which reduces the equations ($9$)-($11$) to,

\begin{align}
(\mathrm{ln}a)^{\prime \prime }+a^{2}c^{2}& =\frac{1}{3}\rho _{r}a^{4}c^{2},
\label{a1} \\
(\mathrm{ln}c)^{\prime \prime }& =\frac{1}{3}\rho _{r}a^{4}c^{2},  \label{a2}
\\
2(\mathrm{ln}a)^{\prime }(\mathrm{ln}c)^{\prime }+(\mathrm{ln}a)^{\prime 2}&
=-a^{2}c^{2}+\rho _{r}a^{4}c^{2}.  \label{a3}
\end{align}

We find the parametric solution quoted in \cite{DLN}, as follows. Defining $%
\omega $, the ratio of the two terms on the right-hand side of ($14$) by 
\begin{equation}
\omega ^{2}\equiv \frac{a^{2}c^{2}}{3\rho _{r}a^{4}c^{2}},  \label{omega_eq}
\end{equation}
we can express the radiation density in terms of the scale factors using $%
\rho _{r}=C_{r}(a^{2}c)^{-4/3},$ and find $\omega $ in terms of the scale
factors as $3\omega ^{2}=a^{2/3}c^{4/3}/C_{r}$. Hence, using ($12$)-($14$),
we can write 
\begin{equation}
2\left( \mathrm{ln}\,\omega \right) ^{\prime \prime }=\frac{2}{3}\left( \rho
_{r}a^{4}c^{2}-a^{2}c^{2}\right) =\frac{2}{3}\rho _{r}a^{4}c^{2}\left(
1-3\omega ^{2}\right) .
\end{equation}%
Inspecting the above equation we see that the $\rho _{r}a^{4}c^{2}$ term
must be a function of the form $f(\omega, \omega ^{\prime}, \omega ^{\prime
\prime})$. Our next task is to determine the form of this function, so that
once we substitute this form in, we can get a self consistent solution for
the evolution equation for the parameter $\omega$, both by following the
route of using the equation($15$) to obtain $\omega$ in terms of the scale
factors and then using the equations of motion of the scale factors; and
also by using the functional form in terms of the parameter $\omega$ that we
choose for $\rho _{r}a^{4}c^{2}$ in its own evolution equation. After a few
simple trials, it luckily turns out that the ansatz $\rho
_{r}a^{4}c^{2}=\alpha \omega ^{\prime }$ gives us the self consistent
solution we desire, by following both of the routes we described. We shall
demonstrate that this is in fact the case. We choose the ansatz $\rho
_{r}a^{4}c^{2}=\alpha \omega ^{\prime }$ and absorb the constant of
integration $C_{r}$ into the constant $\alpha $. Then we have, 
\begin{equation}
\left( \mathrm{ln}\,\omega \right) ^{\prime }=\frac{1}{3}\alpha (\omega
_{0}+\omega -\omega ^{3}).  \label{omegaeq}
\end{equation}%
where $\omega _0$ is a constant of integration. We now examine the equation
we get for $\rho _{r}a^{4}c^{2}$ by substituting in the equations of motion
This is, 
\begin{equation}
(\mathrm{ln}\,\rho _{r}a^{4}c^{2})^{\prime \prime }=\frac{2}{3}\rho
_{r}a^{4}c^{2}(1-6\omega ^{2}).
\end{equation}%
Using our ansatz, $\rho _{r}a^{4}c^{2}=\alpha \omega ^{\prime }$, we also
find that 
\begin{equation}
(\mathrm{ln}\,\alpha \omega ^{\prime })^{\prime \prime }=\frac{2}{3}\alpha
\omega ^{\prime }(1-6\omega ^{2}).
\end{equation}%
We can write the left hand side of this equation as $\left(\mathrm{ln}%
\,\alpha+\mathrm{ln}\,\omega^{\prime}\right)^{\prime \prime}=(\mathrm{ln}%
\,\omega ^{\prime})^{\prime \prime}$ as $\mathrm{ln}\,\alpha$ is a constant.
We can integrate the above equation once as, 
\begin{equation}
(\mathrm{ln}\,\omega ^{\prime})^{\prime}=\frac{2}{3}\left(\omega _0 -2
\omega ^3+\omega\right)
\end{equation}%
\label{originaleq} We can write out the left hand side of this equation as
follows, 
\begin{equation}
(\mathrm{ln}\,\omega ^{\prime})^{\prime}=\omega ^{\prime}\frac{d}{d\omega}%
\left(\mathrm{ln}(\omega ^{\prime}/\omega)+\mathrm{ln}\omega\right)
\end{equation}
Differentiating and cancelling factors of $\omega ^{\prime}$ from the first
term in the brackets in the above equation, and then substituting in the
right hand side of equation ($17$) for $(\mathrm{ln}\, \omega)^{\prime}$, we
get, 
\begin{equation}
(\mathrm{ln\,}\omega ^{\prime})^{\prime}=(\mathrm{ln}\omega )^{\prime}+\frac{%
1}{3}\alpha\omega(1-3\omega ^2)
\end{equation}
Using the expression we found previously for $(\mathrm{ln}\,\omega)^{\prime}$%
, in equation ($17$), we recover the same right hand side as equation ($21$)
up to some additive integration constants, confirming that our ansatz gives
us a self-consistent solution. Hence, we can write the evolution equation
for the parameter $\omega $ as, 
\begin{equation}
\left( \mathrm{ln}\,\omega \right) ^{\prime }=Q^{1/2}(\omega _{0}+\omega
-\omega ^{3}),
\end{equation}%
where we have redefined our constant $\alpha $ to be the constant $3Q^{1/2}$
for notational consistency with ref.\ \cite{Luk}. In conclusion, we have the
following radiation-era solution in terms of the parameter $\omega $ as in 
\cite{Luk}: 
\begin{equation}
a(\tau )=3^{1/2}Q(\omega _{0}+\omega -\omega ^{3}),  \label{radparametera}
\end{equation}%
and 
\begin{equation}
c(\tau )=\frac{3^{1/2}\omega ^{3/2}}{Q^{1/2}(\omega _{0}+\omega -\omega
^{3})^{1/2}}.  \label{radparameterc}
\end{equation}

By inspecting the solution, from the evolution equation of the parameter $%
\omega $, that is ($17$), we find that, for $\omega =-1,0,1$, the equation
yields a simple form, 
\begin{equation}
(\mathrm{ln}\;\omega )^{\prime \ }=Q^{1/2}\omega _{0}.
\end{equation}%
However, for the special choices of $\omega =-1,0$, the scale factor $c(\tau
)$ becomes imaginary and zero, respectively. On computing the volume maxima
in terms of $\omega $ for all positive values of $\omega $ numerically, we
find that the value of $\omega $ at which the volume maximum occurs is very
close to $1$. Thus, we conclude that $\omega \sim 1$ is the point marking
the volume maximum as well as the endpoint of the validity of this
parametric solution. \footnote{%
Well before the volume maximum is approached the effects of the anisotropic
3-curvature are similar to the addition of a trace-free anisotropic pressure
stress (long-wavelength homogeneous gravitational wave modes) on a
background of simple Bianchi I form \cite{skew} or Friedmann form \cite%
{yudin}. In the presence of isotropic black body radiation the two scale
factors evolve as $a(t)\propto t^{1/2}(\ln t)^{2n_{1}}$ and $c(t)\propto
t^{1/2}(\ln t)^{n_{2}}$, with $2n_{1}+n_{2}=0$, so the volume $a^{2}b\propto
t^{3/2}$, evolves as in Friedmann, to leading order \cite{DLN, skew} but the
shear falls more slowly than when the 3-curvature is isotropic.}

We now use the axisymmetric solution for the Type IX universe to see how the
anisotropy behaves at the maximum of each cycle as we increase the radiation
entropy, which is proportional to $C_{r}^{3/4}$. Keeping in mind that the
limit $\omega \longrightarrow 1$ corresponds to the instant of maximum
volume of the expansion, we can see how the ratios of the scale factors in
the two directions behave in this limit. In the limit of $\omega
\longrightarrow 1$, the ratio $a/c$ reduces to, 
\begin{equation}
\frac{a^{2}}{c^{2}}=Q^{3}\omega _{0}^{3}.
\end{equation}%
We have seen that the quantity $Q$ is related to the constant $\alpha ,$ and
hence to the constant $C_{r}$ (which has been absorbed into the constant $%
\alpha $ as stated above). Therefore an increase in the entropy of radiation
(note that equation ($8$) ensures that $dS/dt>0$ if and only if $dS/d\tau >0$%
) causes a corresponding increase in the ratio of the scale factors, which
indicates that the universe is becoming more anisotropic as the heights of
the successive expansion maxima increase.

\subsection{Adding a positive cosmological constant}

We can use a similar approximation, but this time in comoving proper time, $%
t,$ instead of the conformal time coordinate, $\tau $. Using the
approximation that $c\ll a=b$, we can reduce the Einstein equations to the
following form:

\begin{align}
2\frac{\ddot{a}}{a}+\frac{\dot{a}^{2}}{a^{2}}+\frac{1}{a^{2}}& =\Lambda , \\
\frac{\ddot{c}}{c}+\frac{\ddot{a}}{a}+\frac{\dot{c}}{c}\frac{\dot{a}}{a}&
=\Lambda ,  \notag
\end{align}%
where $\Lambda $ is the cosmological constant. If we rewrite the first
equation in the form 
\begin{equation}
a\frac{d}{da}\left( \dot{a}^{2}\right) +\dot{a}^{2}=\Lambda a^{2}-1,
\end{equation}%
it integrates to,

\begin{equation}
\dot{a}^{2}=\frac{\Lambda a^{2}}{3}-1+\frac{C_{1}}{a},  \label{frlambda}
\end{equation}%
where $C_{1}$ is an integration constant. In order to determine the value of
the constant $C_{1}$, we can choose Kasner initial conditions where the
scale factor $a(t)\sim t^{2/3}$ as $t\rightarrow 0$. Substituting this into
equation ($27$), we get $C_{1}=4/9$ as $t\rightarrow 0$. The above equation
resembles the equation for ordinary dust and a cosmological constant in an
isotropic closed Friedmann universe. In the case of cosmological constant
domination, the solution for $a(t)$ tends to the de Sitter solution, that
is, 
\begin{equation}
a(t)\sim \exp \left( \sqrt{\frac{\Lambda }{3}}t\right)
\end{equation}

Under the approximation $a\sim b\gg c$, the Friedmann constraint becomes, 
\begin{equation}
\frac{\dot{a}^{2}}{a^{2}}+2\frac{\dot{a}}{a}\frac{\dot{c}}{c}+\frac{1}{a^{2}}%
=\Lambda  \label{Friedmann_constraint_with_approx}
\end{equation}%
Substituting in ($27$), we get, 
\begin{equation}
2\frac{\dot{a}}{a}\frac{\dot{c}}{c}+\frac{C_{1}}{a^{3}}=\frac{2\Lambda }{3}
\end{equation}%
Substituting in the solution for the cosmological constant dominated $a(t)$,
and taking the limit of very late times (which is the limit in which we
expect cosmological constant domination), we get 
\begin{equation}
c(t)\sim \exp \left( \sqrt{\frac{\Lambda }{3}}t\right)
\end{equation}

Thus we see that for cosmological constant domination isotropisation is
achieved as $t\rightarrow \infty ,$ with the scale factors evolving towards,

\begin{equation}
a(t)=b(t)\sim c(t)\sim \exp \left( \sqrt{\frac{\Lambda }{3}}t\right)
\label{des}
\end{equation}

It is worth noting that this result is not just a case of the standard
cosmic no-hair theorem for spatially homogeneous universes due to Wald \cite%
{wald} because the three-curvature can be positive in type IX universes and
all cosmological no-hair theorems assume that $^{3}R\leq 0$, \cite{jenst}, 
\cite{bgotz}, \cite{jbnoh}. Ostensibly, this is to ensure the universe does
not suffer collapse to a future singularity before the $\Lambda $ term can
dominate. However the Kantowski-Sachs spatially homogeneous universe has $%
^{3}R>0$ and need not approach the de Sitter metric at large $t$ when $%
\Lambda >0$, \cite{KS}. In fact, the conditions necessary (and sufficient)
for type IX models with $\Lambda =0$ to recollapse are extremely subtle \cite%
{bt, bgt} and examples have been found where type IX universes expand
forever even though the sum of the density and the three principal pressures
is positive \cite{hcal}. Typically, the 3-curvature is \textit{negative} as
long as the dynamics are significantly anisotropic ($%
^{3}R=2/c^{2}-b^{2}/2c^{4}$ for axisymmetric \ref{metric} when $a=b\gg c$).
This causes the expansion to continue until there is sufficient
isotropisation for the three-curvature to become positive and only then does
an expansion maximum of the volume become possible. This occurs unless a
positive cosmological constant comes to dominate before it is reached, as in
equation ($32$).

Hence, we see that the results of Barrow and Dabrowski \cite{bdab} showing
the inevitable termination of oscillations in a closed oscillation universe
with $\Lambda >0$ continue to hold in Bianchi type IX universes. Eventually,
cycles will grow large enough for the $\Lambda $ term to dominate the
dynamics. When it does so, it quickly stops the scale factors from reaching
expansion maxima. They all continue to expand and the dynamics are
increasingly dominated by the $\Lambda $ term and approach the de Sitter
metric.

\section{\label{sec:numerics}Numerical solutions of the type IX equations}

We move now to consider the numerical integration of the full
(non-axisymmetric) type IX equations; first, in the presence only of
radiation, $\rho _{r}$ and then with radiation and a ghost field, $\rho _{g}$%
, that will create a smooth bounce at a finite volume minimum. In each case
we will be interested in the behaviour with ($\dot{S}>0$) and without ($\dot{%
S}=0$) entropy increase and with and without a cosmological constant. These
computations will enable us to confirm the general picture found from the
analytic approximations of the previous section.

\subsection{Radiation universe with no entropy increase: $\dot{S}=0$}

\subsubsection{No ghost field: $\protect\rho _{g}=0$}

In the case of the Bianchi IX universe containing just the radiation field
we have followed the initial assumption of \cite{DLN} that two of the scale
factors approach the same value in the expanding half of the cycle. Thus on
approach to the maximum of the scale factor, the universe resembles an
axisymmetric type IX universe. The scale factors that are similar to each
other approach their maxima first and then, after reaching their maxima,
they start contracting and then oscillate around an almost constant value.
The third scale factor approaches its maximum at a later time and goes past
the peaks of the other two scale factors before reversing into contraction.
We can now follow the evolution of the quantities in this scenario, without
needing to assume axial symmetry, by computing the behaviour of the scale
factors from various sets of initial conditions.

We start with initial conditions that are similar to the ones chosen in 
\textbf{\ }\cite{DLN}\textbf{,} that is, 
\begin{equation}
\lbrack a(t),b(t),c(t)]\propto \lbrack t^{1/2},t^{1/2},t^{5/8}].
\label{initial}
\end{equation}%
In this case, in the absence of the ghost field, we find that the universe
is unable to re-expand after collapsing. The addition of the radiation or
dust field does not cause a qualitative change in the behaviour of the scale
factors, but causes the system to reach stiffness faster. Thus we choose
values for the initial conditions for the density of radiation to be $\rho
_{r}(t_{i})=1$ where $t_{i}$ refers to the initial instant of time at which
we start the integration. The scale factors in two directions oscillate
about each other as they follow the evolution trend of the volume scale
factor. The third scale factor is then smaller than the other two scale
factors and does not display the oscillatory behaviour undergone by the
other two scale factors. The shear and the $3$-curvature show oscillatory
profiles before blowing up on approach to the strong curvature singularity
at the `big crunch'. The presence of the singularity at the end of the
collapse phase is inferred by the fact that the density of the matter and
radiation components diverge there.

Running the simulation with arbitrarily selected initial conditions or even
Kasner-like initial conditions, makes the collapse occur closer to the
starting instant, in comparison to the case done with the initial conditions
in ($33$) and little information can be extracted from the results. For the
Kasner initial conditions, however, before the collapse occurs, the
individual scale factors show some oscillatory behaviour.

\begin{figure}[tbp]
\caption{Time evolution of the volume (top) and three orthogonal scale
factors (bottom) of a type IX universe  with only radiation and no ghost
field in comoving proper time $t$ during a single cycle. The blue dashed,
green dotted and yellow solid lines correspond to scale factors $a(t)$, $b(t)
$ and $c(t)$ respectively.}\centering\hfill \break 
\begin{minipage}{0.5\textwidth}
\includegraphics[width=1.0\linewidth, height=0.3\textheight]{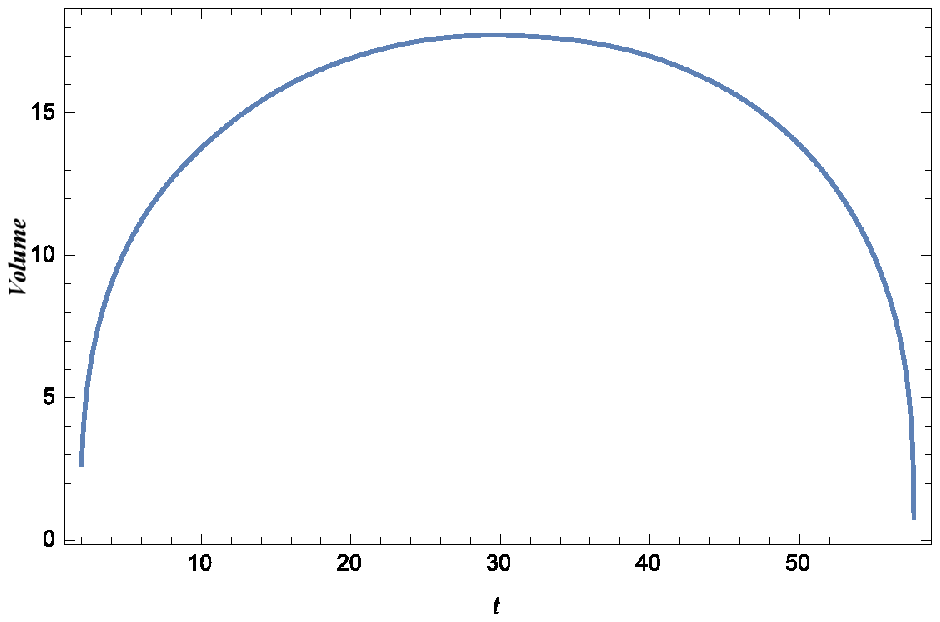}
\subcaption{\label{fig:withoutstifffieldvolume}}
\end{minipage}\hfill 
\begin{minipage}{0.5\textwidth}
\includegraphics[width=1.0\linewidth, height=0.3\textheight]{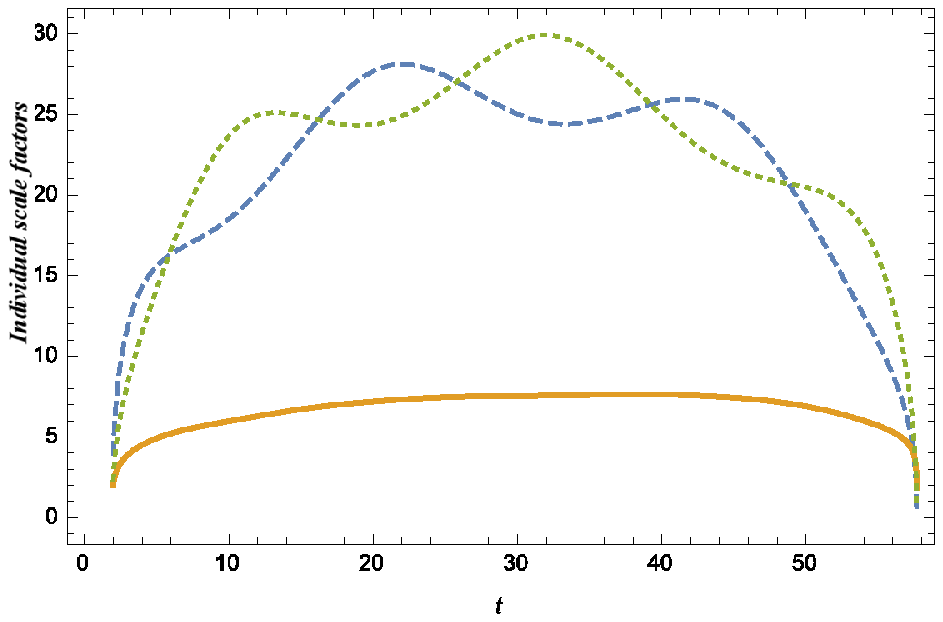}
\subcaption{\label{fig:withoutstifffieldindividual}}
\end{minipage}
\end{figure}

\begin{figure}[tbp]
\caption{From top to bottom: Evolution of shear and $3$-curvature scalars
during a single cycle of a type IX universe containing radiation and no
ghost field, with comoving proper time, $t$.}\centering
\begin{minipage}{0.4\textwidth}
\includegraphics[width=1.0\linewidth, height=0.2\textheight]{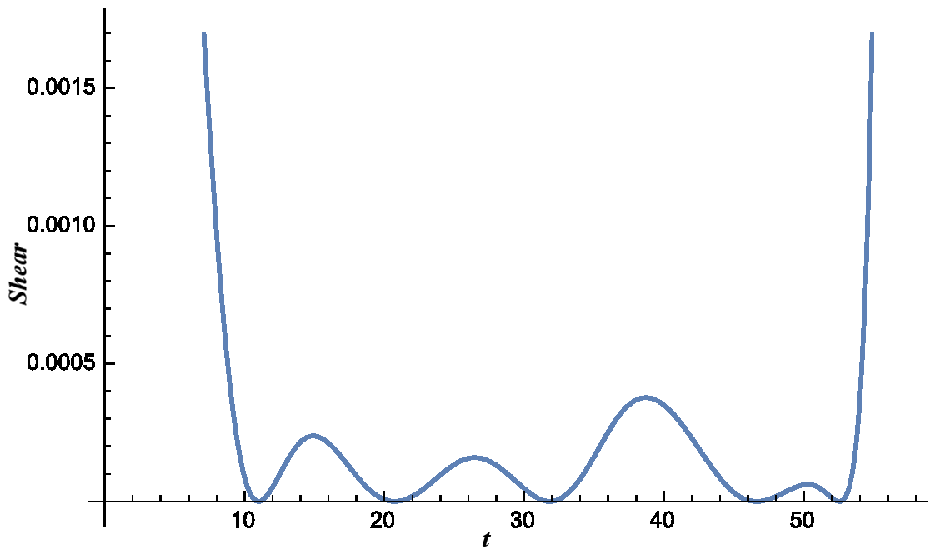}
\subcaption{\label{fig:withoutstifffieldshear}}
\end{minipage}\hfill 
\begin{minipage}{0.5\textwidth}
\includegraphics[width=0.8\linewidth, height=0.2\textheight]{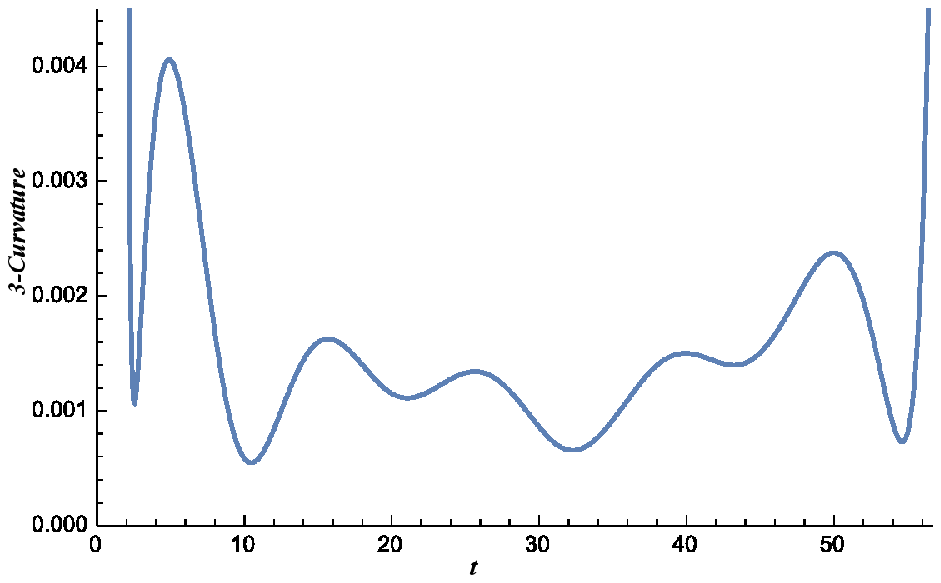}
\subcaption{\label{fig:withoutstifffield3curvature}}
\end{minipage}
\end{figure}

We see a single oscillation of the volume scale factor in Figure \ref%
{fig:withoutstifffieldvolume}. The behaviours of the individual scale
factors are seen in Figure \ref{fig:withoutstifffieldindividual}, where the
blue dashed line corresponds to the scale factor $a(t)$, the green dotted
and the yellow solid lines to the scale factors $b(t)$ and $c(t),$
respectively. As we have noted before, the scale factors $a(t)$ and $b(t)$
show small oscillations around each other, while the scale factor $c(t)$ has
a much smaller amplitude and does not show such oscillations. This is
reminiscent of a long era in the evolution of type IX on approach to an
initial singularity seen in ref \cite{bkl}. On looking at the shear and the
spatial $3$-curvature, we see that they display oscillatory behaviour as
well, before blowing up on approach to the singularity. The shear evolution
is shown in Figure \ref{fig:withoutstifffieldshear} and the $3$-curvature
evolution is shown in Figure \ref{fig:withoutstifffield3curvature}.

\subsubsection{Ghost field present: $\protect\rho _{g}\neq 0$}

We create a simple bouncing cosmological model by adding a ghost field to
create a non-singular bounce. The other fields in the system are the
radiation fields and the dust fields. As before, we again choose initial
values for the radiation, the dust and the `ghost' fields to be of order $1$%
, as $\rho _{r}(t_{i})=8$, $\rho _{m}(t_{i})=5$ and $\rho _{g}(t_{i})=-5$,
respectively. Again, as long as the initial conditions for the densities are
of the same order, their exact numerical value does not much affect the
results of the computation qualitatively. Changing these numbers
significantly changes the number of bounces the system undergoes in the same
time frame of integration but qualitatively the features do not alter. Of
course, as one might expect, changing the initial conditions for the density
of the `ghost' field to be orders of magnitude smaller than the radiation
and dust fields, causes the system to collapse and not undergo the
non-singular bounce. On evolving this system through several successions of
bounces, we find that the scale factors oscillate rapidly from cycle to
cycle as do the energy densities of the radiation, matter and ghost field.
The square of the shear tensor and the $3$-curvature also show similar
oscillatory behaviour.

\begin{figure}[tbp]
\caption{Evolution of volume (top) and individual scale factors (bottom) with radiation,
ordinary dust and the ghost field included, with time.The blue dashed, green
dotted and yellow solid lines correspond to $a(t)$, $b(t)$ and $c(t)$
respectively.}\centering\hfill \break 
\begin{minipage}{0.48\textwidth}
\includegraphics[width=1.0\linewidth, height=0.3\textheight]{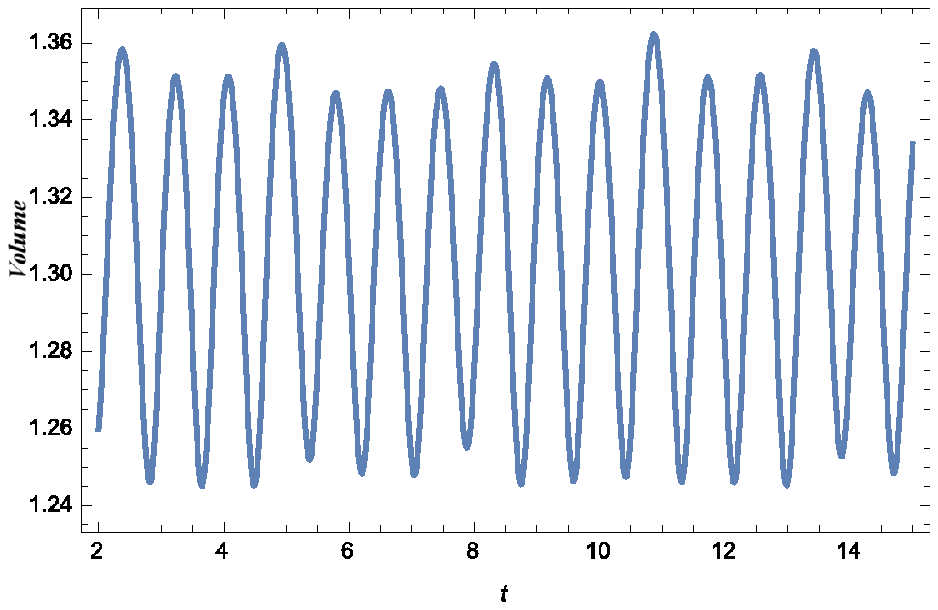}
\subcaption{\label{fig:oscillationswithghostvolume}}
\end{minipage}\hfill 
\begin{minipage}{0.48\textwidth}
\includegraphics[width=1.0\linewidth, height=0.3\textheight]{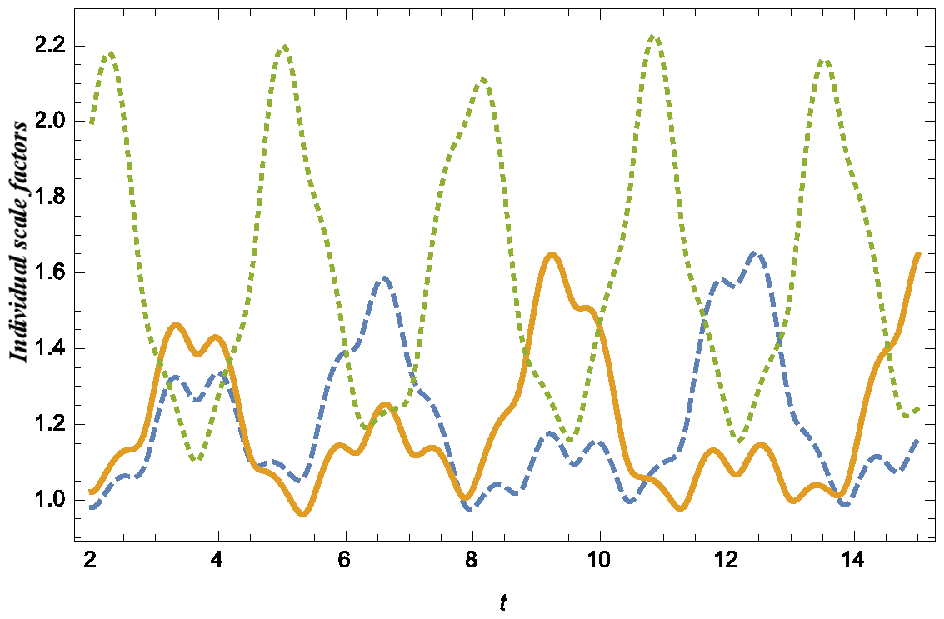}
\subcaption{\label{fig:oscillationswithghostindividual}}
\end{minipage}
\end{figure}

\begin{figure}[tbp]
\caption{From top to bottom: Evolution of shear and curvature with
radiation, ordinary dust and the ghost field included, with time}\centering
\hfill \break 
\begin{minipage}{0.48\textwidth}
\includegraphics[width=1.0\linewidth, height=0.3\textheight]{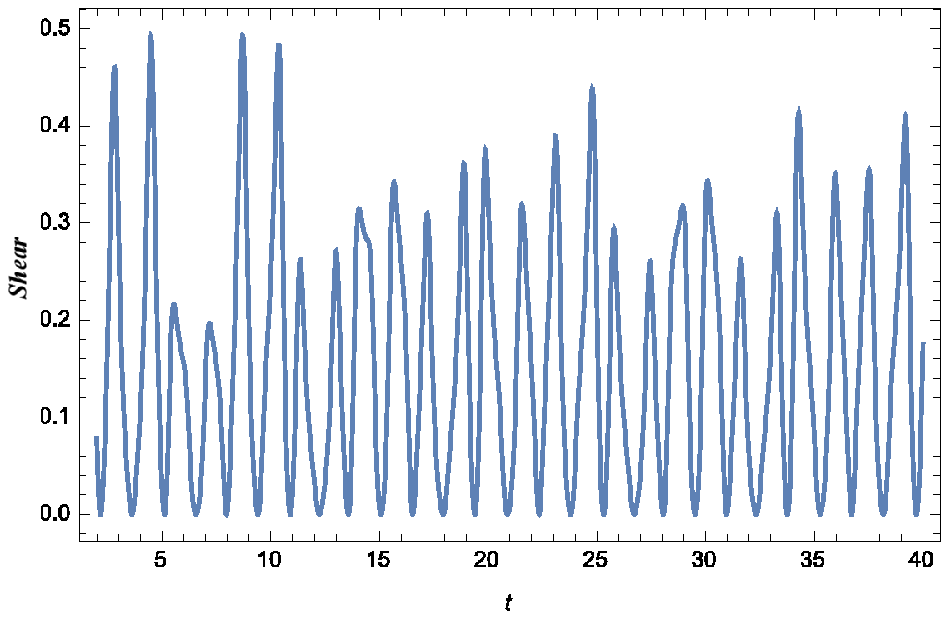}
\subcaption{\label{fig:oscillationswithghostshear}}
\end{minipage}\hfill 
\begin{minipage}{0.48\textwidth}
\includegraphics[width=1.0\linewidth, height=0.3\textheight]{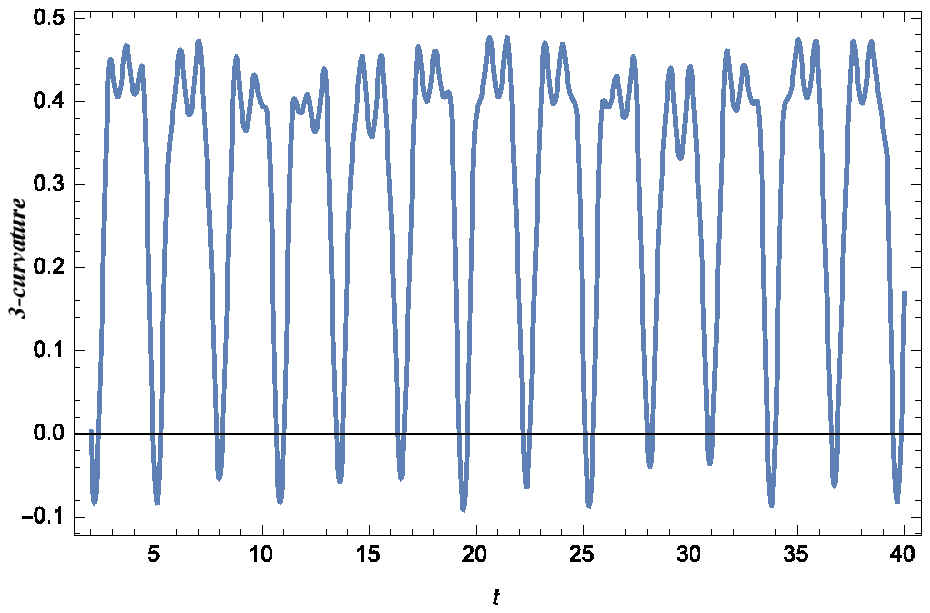}
\subcaption{\label{fig:oscillationswithghostcurvature}}
\end{minipage}
\end{figure}

The evolution of the volume scale factor from cycle to cycle is oscillatory,
as can be seen in Figure \ref{fig:oscillationswithghostvolume}, as are the
behaviors of the scale factors in the three orthogonal directions. As
before, in Figure \ref{fig:oscillationswithghostindividual}, the blue
dashed, solid yellow, and dotted green lines trace the $a(t)$, $b(t)$ and $%
c(t)$ scale factors. The shear and the $3$-curvature also display
oscillatory behaviour and are shown in Figures \ref%
{fig:oscillationswithghostshear} and \ref{fig:oscillationswithghostcurvature}
respectively.

Therefore, we see that adding the ghost field is essential to avoid collapse
to a singularity after just one cycle, and to allow it to actually propagate
smoothly through successive bounces. We will include the ghost field to the
model in the rest of the paper when we examine the effects of entropy
injection, or the effects of adding a cosmological constant, so that we can
evolve the model through a series of cycles without encountering
singularities.

\subsection{Radiation universe with entropy increase: $\dot{S}>0$}

\subsubsection{Ghost field present: $\protect\rho _{g}\neq 0$}

We now consider our bouncing anisotropic cosmological model with dust,
radiation and a ghost field to prompt and to allow it to propagate through
several cycles when there is entropy increase from cycle to cycle.

We start with Kasner initial conditions, and with the same initial
scale-factor evolution equation ($33$) and the respective energy densities
as before($\rho _{r}(t_{i})=8$, $\rho _{m}(t_{i})=5$ and $\rho
_{g}(t_{i})=-5 $). This time, we increase the value of the constant of
radiation ($C_{r}$) by a factor of $2$, to model the effects of entropy
increase on the dynamics. We find that the volume scale factor shows an
increase in cycle-to-cycle expansion maxima as expected. The individual
scale factors proceed through several chaotic oscillations in each cycle,
and the three directions seem to oscillate increasingly out of phase as the
volume maxima get larger.

\begin{figure}[tbp]
\caption{Evolution of the volume (top) and individual scale factors (bottom) of a type IX
universe with entropy increase from cycle to cycle ($dS/dt>0$), with $t$
time. The blue dashed, green dotted, and solid yellow lines correspond to
the orthogonal scale factors $a(t)$, $b(t)$ and $c(t)$ respectively. The
model includes radiation, dust matter and the ghost field. The ghost field
ensures that the cycles bounce smoothly at finite values of the volume.}%
\centering\hfill \break 
\begin{minipage}{0.48\textwidth}
\includegraphics[width=1.0\linewidth, height=0.3\textheight]{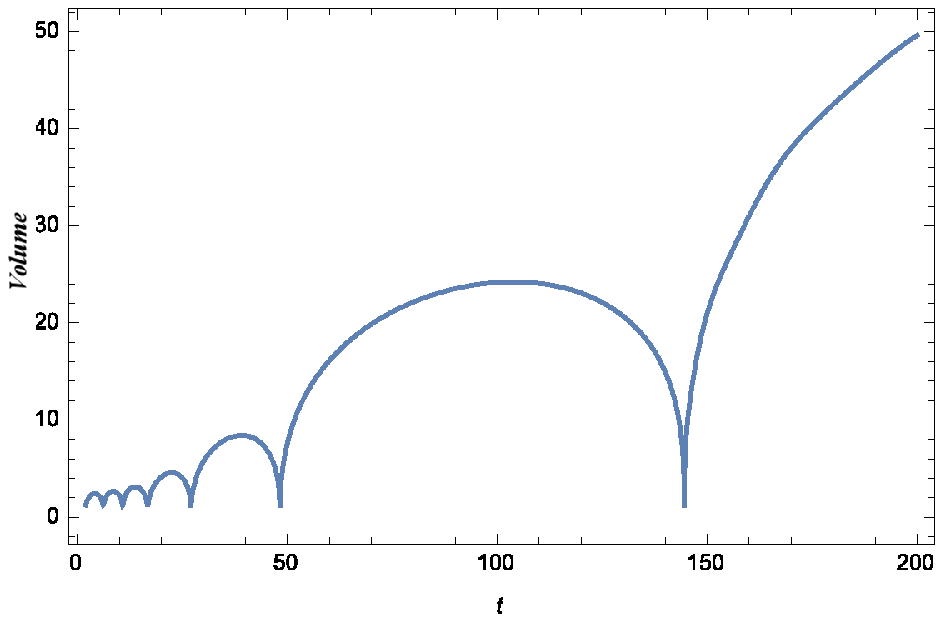}
\subcaption{\label{fig:entropyvolume}}
\end{minipage}\hfill 
\begin{minipage}{0.48\textwidth}
\includegraphics[width=1.0\linewidth, height=0.3\textheight]{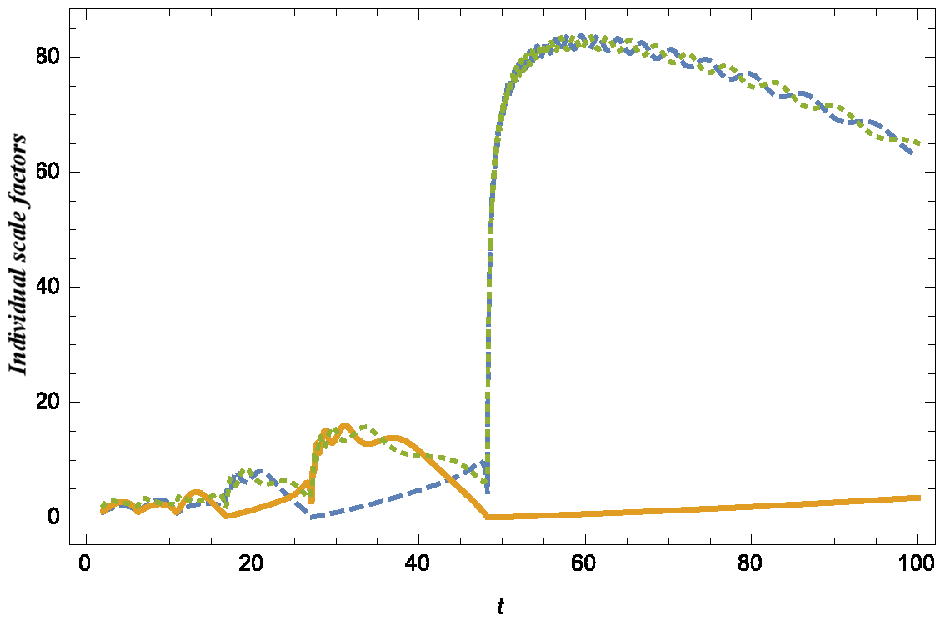}
\subcaption{\label{fig:entropyindividual}}
\end{minipage}
\end{figure}

Figure \ref{fig:entropyvolume} represents the evolution of the volume scale
factor and Figure \ref{fig:entropyindividual} represents the evolution of
the individual scale factors.

To see if greater expansion-volume maxima lead to an increase in the
anisotropy, we plot the square of the shear tensor(Figure \ref%
{fig:entropyshear}), denoted by $\sigma ^{2}$, and we see that the shear
tensor indeed shoots up to larger and larger values, at each successive
minimum as the corresponding radiation maximum is increased. We can see that
a similar increase occurs when we track the difference in the expansion
rates of the scale factors in the three directions. A significant increase
in the differences of the expansion rates in the $a$ and the $b$ directions,
and in the difference of the expansion rates in the $b$ and $c$ directions
is seen as the expansion maxima get bigger. There is not such a large
increase in the difference in the expansion rates in the $a$ and $c$
directions. We can also look at the $3$-curvature(Figure \ref%
{fig:entropycurvature}), and we find that it oscillates to a greater extent
initially, but as time increases, the amplitude and frequency of these
oscillations decrease, and the universe seems to approach flatness, albeit
with strong expansion and $3$-curvature anisotropy.

\begin{figure}[tbp]
\caption{ Evolution of shear (top) and the $3$-curvature (bottom) scalars with time
in a type IX universe with entropy increase. The model includes radiation,
dust and the ghost field to create smooth non-singular bounces.}\centering%
\hfill \break 
\begin{minipage}{0.48\textwidth}
\includegraphics[width=1.0\linewidth, height=0.3\textheight]{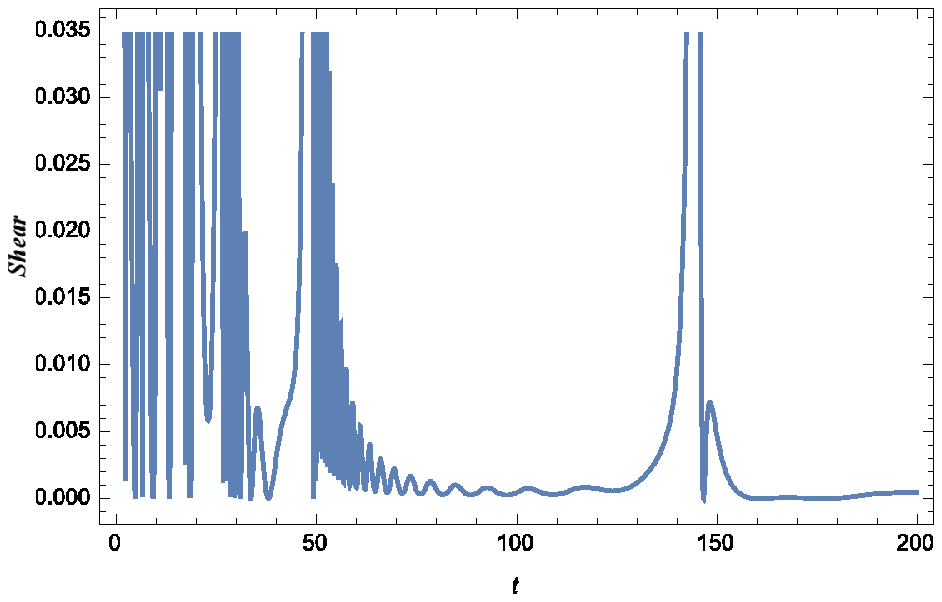}
\subcaption{\label{fig:entropyshear}}
\end{minipage}\hfill 
\begin{minipage}{0.48\textwidth}
\includegraphics[width=1.0\linewidth, height=0.3\textheight]{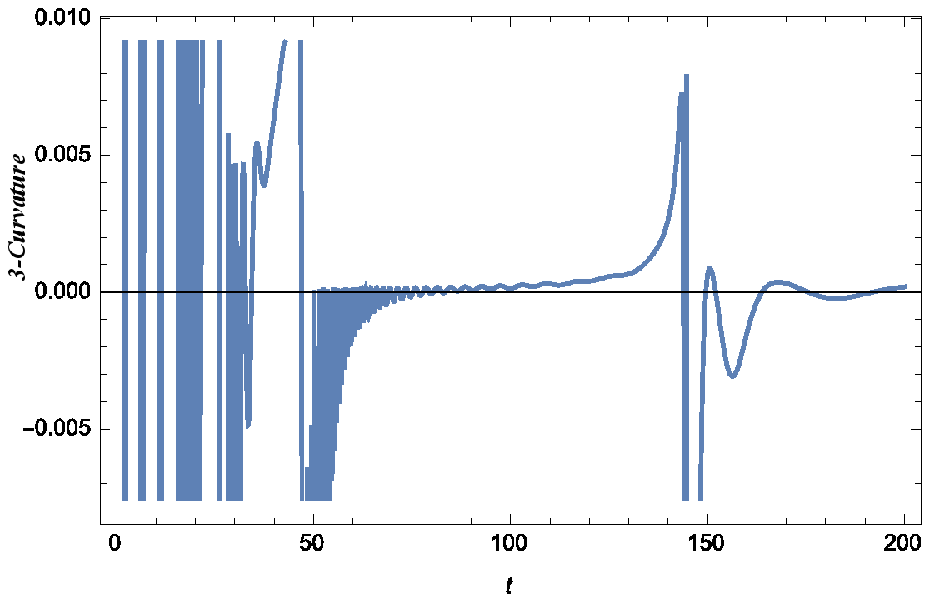}
\subcaption{\label{fig:entropycurvature}}
\end{minipage}
\end{figure}

We can understand this intuitively in another way. The shear is at its
highest near the minimum of each cycle. As the expanding phase of the cycle
begins, the shear is diluted rather slowly as $\sigma \sim (t\mathrm{ln}%
t)^{-1}$. Until the shear is diluted sufficiently, the universe cannot
recollapse. When the shear enters the contracting phase, shear anisotropy
accumulates. The longer the model evolves before the bounce, the greater
amount of shear anisotropy is accumulated. Thus, by injecting radiation
entropy, and effectively increasing the expansion maximum, we give the shear
anisotropy more time to increase in the contracting phase, and thus, more
time to dilute in the expanding phase. Hence, despite the increase in the
shear anisotropy, we still see the universe recollapse and bounce.

We can also compare the pattern of entropy growth in the anisotropic Bianchi
IX case with its isotropic subcase, the closed Friedmann universe.

\begin{figure}[tbp]
\centering
\hfill \break 
\begin{minipage}{0.48\textwidth}
\includegraphics[width=1.0\linewidth, height=0.3\textheight]{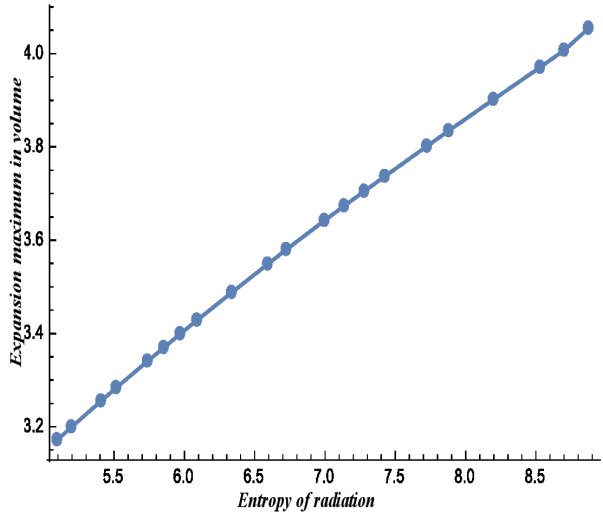}
\subcaption{\label{fig:volentropybianchi}}
\end{minipage}\hfill 
\begin{minipage}{0.48\textwidth}
\includegraphics[width=1.0\linewidth, height=0.3\textheight]{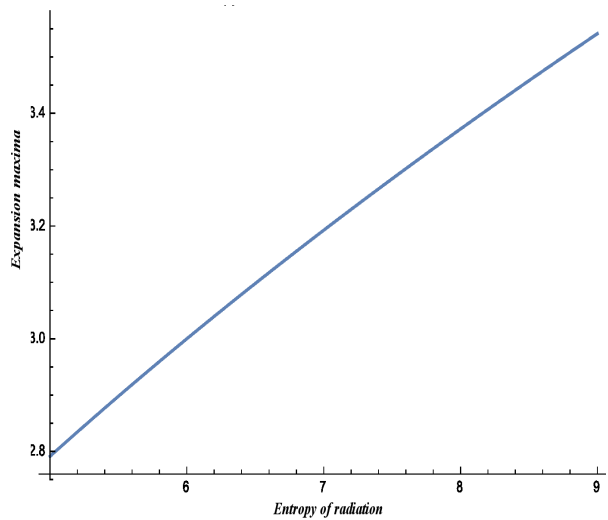}
\subcaption{\label{fig:volentropyFRW}}
\end{minipage}
\caption{(Top) Volume maxima plotted against entropy of radiation in Bianchi
IX universes and (bottom) in isotropic Friedmann universes.}
\end{figure}

We see in Figure \ref{fig:volentropybianchi} for the case of the Bianchi IX
universe and in Figure \ref{fig:volentropyFRW} for the case of the isotropic
closed Friedmann universe, that the variation of successive volume scale
factor maxima in the Friedmann and the type IX universes with increasing
radiation entropy show very similar, almost linear behaviour (this would be
different if we injected entropy according to a different rule). We also
show what happens to the range of the bounce with entropy injection in the
Bianchi IX case.

\begin{figure}[tbp]
{H} \centering
\includegraphics[width=1.0\linewidth,height=0.3%
\textheight]{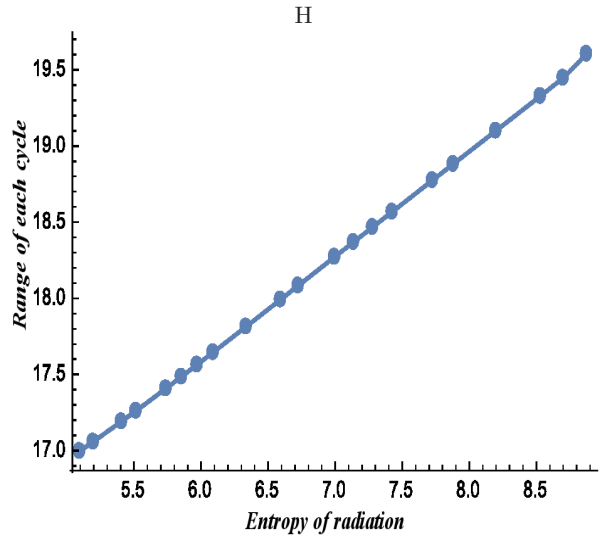} 
\caption{Range of bounces versus entropy of radiation in the presence of
ordinary dust and radiation in oscillating Bianchi IX universes.} \label%
{fig:rangeentropybianchiix}
\end{figure}

It should come as no surprise that the range is also increasing fairly
linearly with the injection of entropy as can be seen in Figure \ref%
{fig:rangeentropybianchiix}. The increase in volume maxima, simply means
that the model takes a longer time to recollapse.

\section{Adding a cosmological constant}

We have seen in previous work and that the addition of cosmological constant
to the closed Friedmann model which incorporates increasing volume maxima
with the injection of radiation entropy, results in the model ceasing to
oscillate before expanding exponentially towards the de Sitter metric \cite%
{bdab}. Now we study the effects of adding both a positive or a negative
cosmological constant in type IX universes. The motivation for doing this,
for the case of the positive cosmological constant, is to see if a similar
exponential expansion to the isotropic case takes place. It is also
interesting to investigate the effect the expansion prompted by the
cosmological constant has on the anisotropy and the spatial 3-curvature. The
negative cosmological constant models create an interesting anisotropically
recollapsing counterpart to the closed Friedmann models. These models are
among the simplest versions of a closed isotropic bouncing universe as they
incorporate a negative cosmological constant as well as a curvature `field'
which behaves as a fluid with equation of state $p=-1/3\rho $. They admit a
simple periodic solution in the isotropic case. It is expected that when
anisotropy is included, in the absence of an ultra stiff matter field, the
universe will quickly approach an anisotropic singularity since the negative 
$\Lambda $ term is only influential at large volumes to effect a collapse
but has negligible effect as the future singularity is reached.

\subsection{Positive cosmological constant}

We study the effect of adding a positive cosmological constant to a type IX
model containing radiation with entropy injection, a dust field, and the
ultra-stiff ($p=5\rho $) ghost field to ensure a smooth non-singular bounce.
The results depend on the value of the cosmological constant relative to the
initial energy densities of the other fields. If the cosmological constant
is set to have a value of the same order of magnitude as the other
components in the system, then the model displays a very sudden increase in
all the scale factors in each of the three directions without any of the
oscillating behaviour that we have seen in the other cases: it quickly
asymptotes to de Sitter behaviour. On the other hand, if the cosmological
constant is too small then its effects cannot be seen in the time interval
that we have set for integration. Thus, for an intermediate range of values
of the cosmological constant, we find that as soon as the cosmological
constant starts to dominate, the volume scale factor which was hitherto
showing an oscillatory behaviour with increasing maxima from cycle to cycle,
does not recollapse beyond the point of maximum. Instead, it enters a final
phase of exponential expansion. The individual scale factors all undergo
exponential expansion, asymptoting to de Sitter behaviour. However, they
have different rates of expansion in this phase, with two of the scale
factors having nearly the same rate, oscillating around each other in this
phase (this is reminiscent of the axisymmetric behaviour that motivated our
analytic approximation in section III above). The third scale factor expands
much less than the other two. As soon as the cosmological constant starts to
dominate, the shear and the 3-curvature, which were oscillating before $%
\Lambda $-domination as in the previous case with $\Lambda =0$, now start
oscillating with smaller and smaller amplitudes as time progresses. For the
purposes of our computation, we set the cosmological constant to be a value
which is approximately $\Lambda \approx 3H^{2}$ which can be taken to mark
the onset of cosmological constant domination.

\begin{figure}[tbp]
\caption{Evolution of (top) the volume, and (bottom) the individual scale
factors of a type IX universe with positive $\Lambda $, with time, $t$. The
blue dashed, green dotted and solid yellow lines correspond to scale factors 
$a(t)$, $b(t)$ and $c(t),$ respectively. The model includes radiation, dust
and a ghost field to create non-singular bounces. Note that the oscillations
cease after a finite time when $\Lambda $ term dominates the dynamics at
large volume. All scale factors then asymptote to the de Sitter expansion
after a few transitionary oscillations.}\centering
\hfill \break 
\begin{minipage}{0.48\textwidth}
\includegraphics[width=1.0\linewidth, height=0.3\textheight]{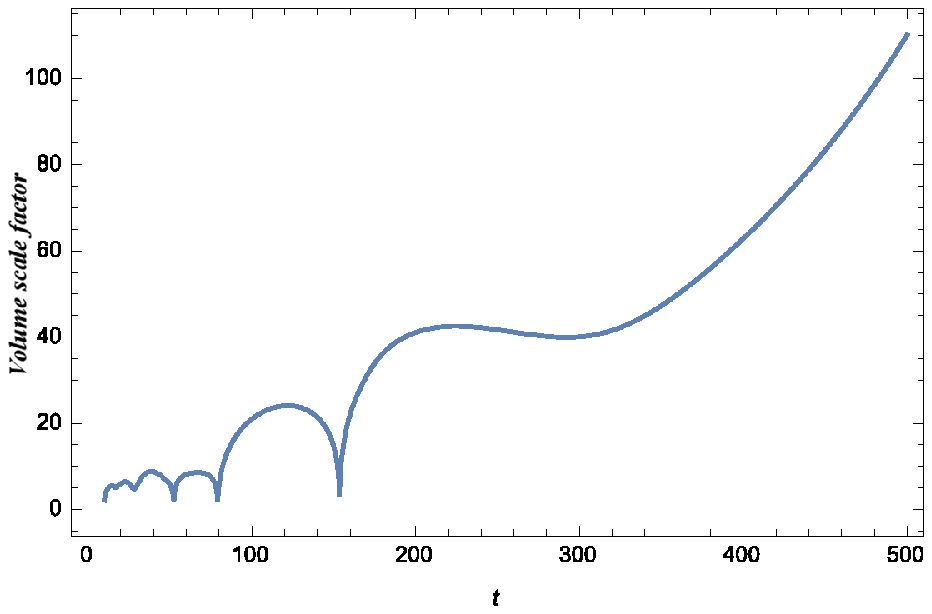}
\subcaption{\label{fig:cosmologicalconstantvolume}}
\end{minipage}\hfill 
\begin{minipage}{0.48\textwidth}
\includegraphics[width=1.0\linewidth, height=0.3\textheight]{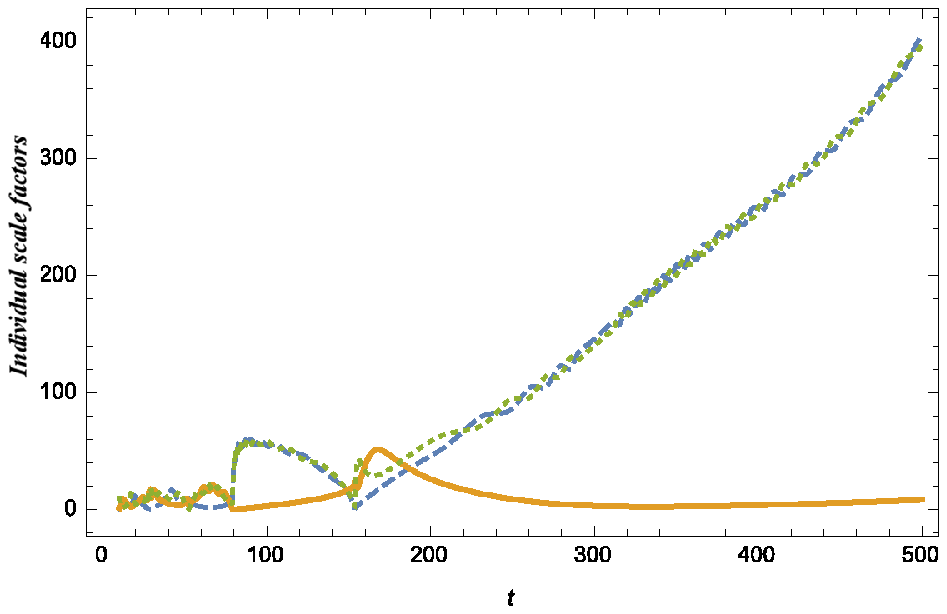}
\subcaption{\label{fig:cosmologicalconstantindividual}}
\end{minipage}
\end{figure}

In Figure \ref{fig:cosmologicalconstantvolume} we show the evolution of the
volume scale factor and, in Figure \ref{fig:cosmologicalconstantindividual},
the evolution of the individual scale factors, where the blue dashed, yellow
solid, and green dotted lines represent the scale factors $a(t)$, $b(t)$ and 
$c(t)$, respectively. We can see the domination of the $\Lambda $ term
leading to de Sitter like expansion in all of the scale factor directions by
looking at the Hubble rates.

\begin{figure}
\caption{Evolution of the Hubble rates in the presence of a positive cosmological constant with time.The
blue dashed, green dotted and solid yellow lines trace the Hubble rates 
$\dot{a}(t)/a(t)$, $\dot{b}(t)/b(t)$ and $\dot{c}(t)/c(t)$. Oscillations cease when $\Lambda$ dominates. The Hubble rates then undergo an anisotropic transition phase before eventually approaching isotropic de Sitter-like expansion where the individual Hubble rates approach the same constant value}
\includegraphics[width=1.0\linewidth,
height=0.3\textheight]{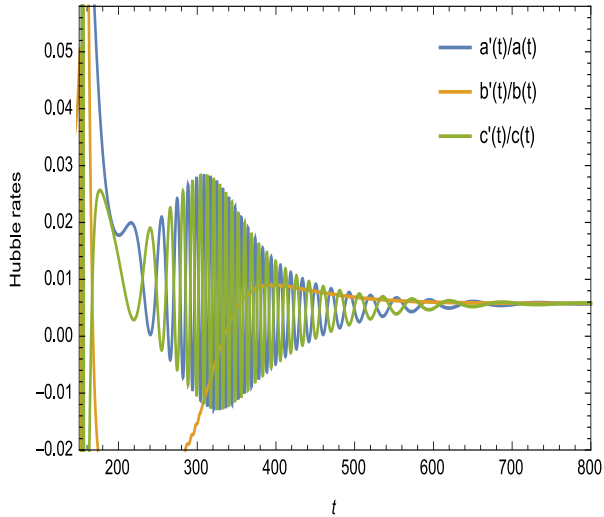} %
\label{fig:cosmologicalconstantindividualHubble}
\end{figure}

We see in Figure \ref{fig:cosmologicalconstantindividualHubble}, that the
Hubble rates undergo several oscillations and enter into a transition phase
when the cosmological constant starts to dominate. However with positive
cosmological constant domination, the oscillations cease, and as the scale
factors undergo exponential expansion, their Hubble rates tend to a constant
value.

\begin{figure}[tbp]
\caption{(Top) Evolution with time, $t$, of the shear and (bottom)  $3$-curvature scalars in type IX universes with positive $\Lambda $. The model
includes radiation, dust and a ghost field to create non-singular bounces.}%
\centering
\hfill \break 
\begin{minipage}{0.48\textwidth}
\includegraphics[width=1.0\linewidth, height=0.3\textheight]{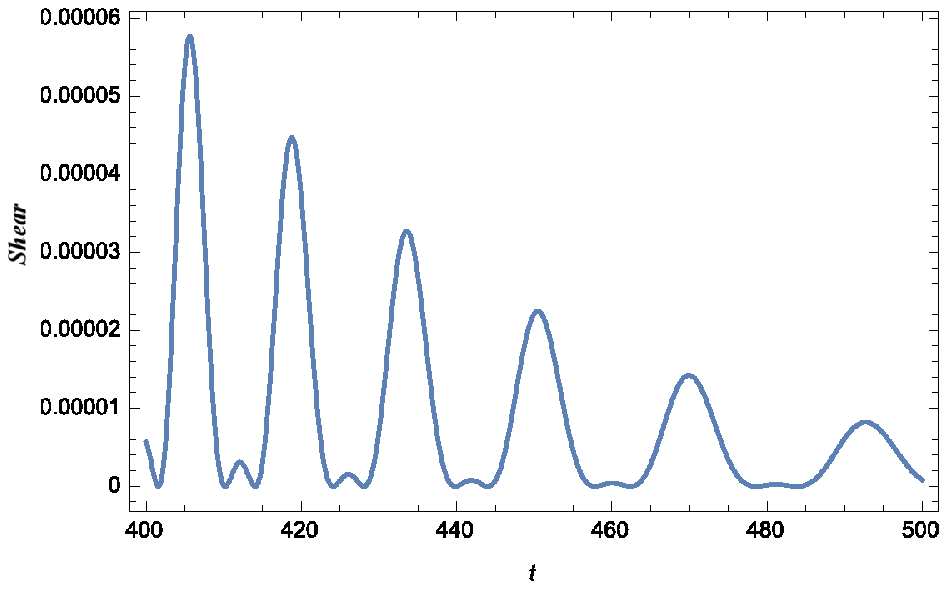}
\subcaption{\label{fig:cosmologicalconstantshear}}
\end{minipage}\hfill 
\begin{minipage}{0.48\textwidth}
\includegraphics[width=1.0\linewidth, height=0.3\textheight]{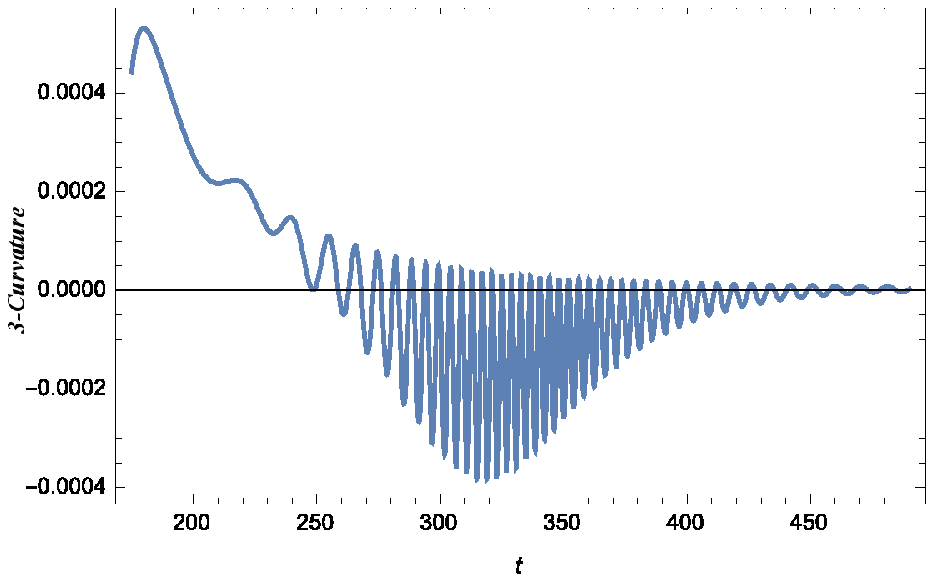}
\subcaption{\label{fig:cosmologicalconstantcurvature}}
\end{minipage}
\end{figure}

Figures \ref{fig:cosmologicalconstantshear} and \ref%
{fig:cosmologicalconstantcurvature} show the behaviour of the shear tensor
squared $\sigma ^2$ and of the $3$-curvature respectively. They show a
decrease with time, with the shear showing oscillations, as before, but with
decreasing amplitude. The curvature also shows oscillations as before but
falls to very small values as the cosmological constant starts to dominate.

\subsection{Negative cosmological constant}

We can also try to look at the case of the negative cosmological constant to
find the role of anisotropies in the simple isotropic universe model. This
is the closed Friedmann universe with curvature parameter $k=+1$, consisting
of `domain wall matter' which is just matter with an equation of state $%
p=-2\rho /3$ $\propto a^{-1}$ and a negative cosmological constant. The
Friedmann equation is now, ($\Sigma >0$, constant): 
\begin{equation}
\frac{\dot{a}^{2}}{a^{2}}=\frac{8\pi G}{3}\left( \Lambda +\frac{\Sigma }{a}%
\right) -\frac{1}{a^{2}}.
\end{equation}%
This model has been studied as a simple model of a bouncing universe
admitting a simple solution (see Figure 12).

\begin{figure}[tbp]
\caption{Evolution in $t$ time of the scale factor for an isotropic closed
Friedmann universe with negative cosmological constant and `domain-wall'
matter ($p=-2\protect\rho /3$). The domain-wall matter acts like a glost
field to produce smooth non-singular bounces. No entropy production is
included.}\includegraphics[width=1.0\linewidth,height=0.2%
\textheight]{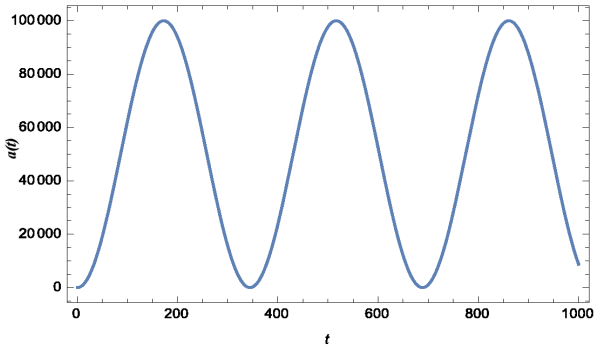}
\end{figure}

Consider an extension of this model to the anisotropic type IX case. In the
absence of a stiff field to smooth out the anisotropies on approach to the
bounce, it is expected that the universe on collapse will not be able to
re-expand from a singularity. To prevent this, we add an ultra stiff matter
field, with positive energy density, as we expect the bounce to be produced
by the negative cosmological constant. We find that the volume scale factor
undergoes a bounce. The individual scale factors undergo several
oscillations with different time periods and different amplitudes. The shear
undergoes oscillations with amplitudes that decrease as the volume scale
factor expands and starts increasing again as the contraction phase begins.
The 3-curvature also shows oscillatory behaviour as the expansion phase is
followed by the contraction.

\begin{figure}[tbp]
\caption{Evolution of the (top) volume and (bottom) individual scale factors
for a type IX universe with negative $\Lambda $, with time, $t$.The blue
dashed, green dotted and yellow solid lines correspond to $a(t)$, $b(t)$ and 
$c(t),$ respectively. The model includes an ultra stiff matter field ($p=5\protect\rho $) with positive energy density in addition to the negative
cosmological constant.}\centering
\hfill \break 
\begin{minipage}{0.48\textwidth}
\includegraphics[width=1.0\linewidth, height=0.3\textheight]{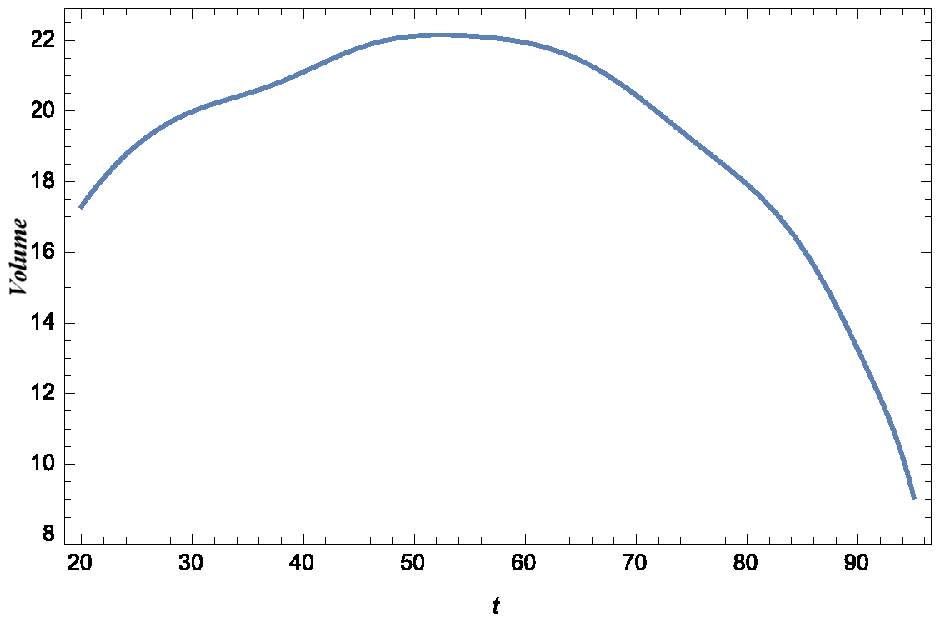}
\subcaption{\label{fig:shuvolume}}
\end{minipage}\hfill 
\begin{minipage}{0.48\textwidth}
\includegraphics[width=1.0\linewidth, height=0.3\textheight]{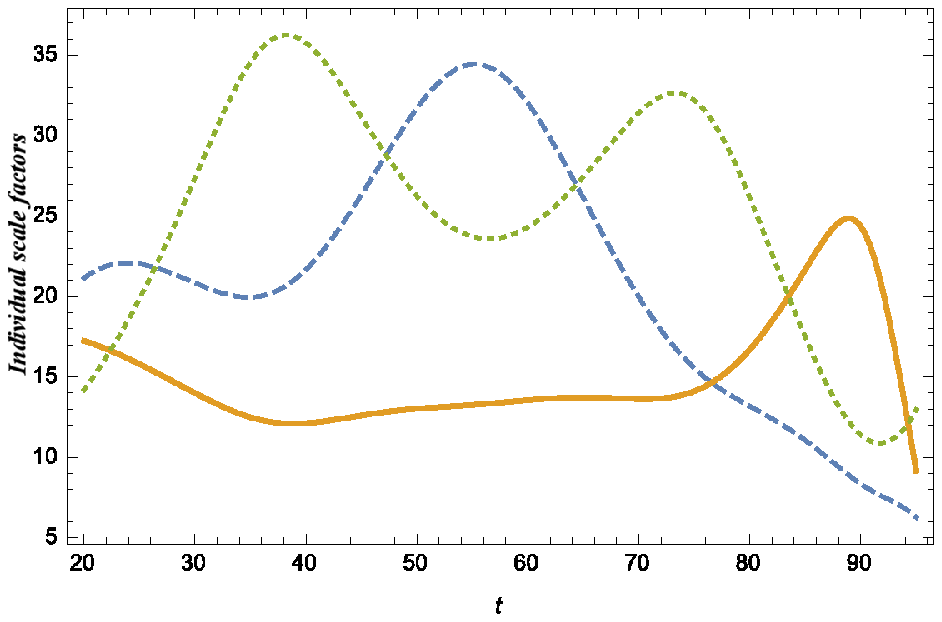}
\subcaption{\label{fig:shuindividual}}
\end{minipage}
\end{figure}

On changing the sign of the cosmological constant in the model, with the
same initial conditions as we have been using previously, we see that with
the injection of entropy into the radiation field, to facilitate an increase
in the expansion maximum, the volume scale factor undergoes irregular
oscillations. These can be seen in Figure \ref{fig:shuvolume}.

The individual scale factors are seen in Figure \ref{fig:shuindividual} and
are given as before by the blue dashed, green dotted and solid yellow lines
representing the scale factors $a(t)$, $b(t)$ and $c(t)$, respectively. They
appear to undergo oscillations as well.

\begin{figure}[tbp]
\caption{(Top) Evolution of the shear and (bottom) the $3$-curvature, with
negative $\Lambda $, versus time, $t$. The model includes an ultra stiff
field ($p=5\protect\rho $) with positive energy density in addition to a
negative cosmological constant.}\centering
\hfill \break 
\begin{minipage}{0.48\textwidth}
\includegraphics[width=1.0\linewidth, height=0.3\textheight]{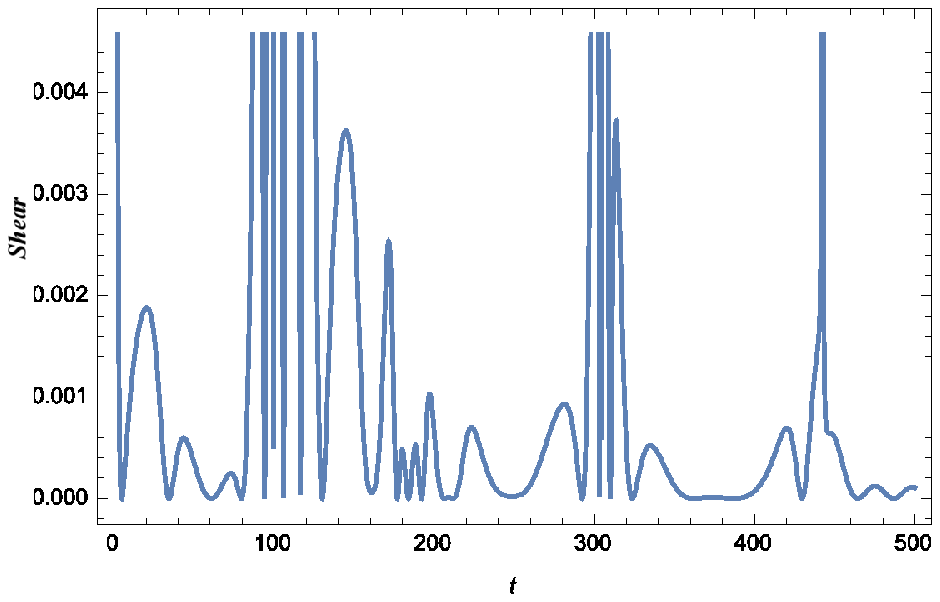}
\subcaption{\label{fig:shushear}}
\end{minipage}\hfill 
\begin{minipage}{0.48\textwidth}
\includegraphics[width=1.0\linewidth, height=0.3\textheight]{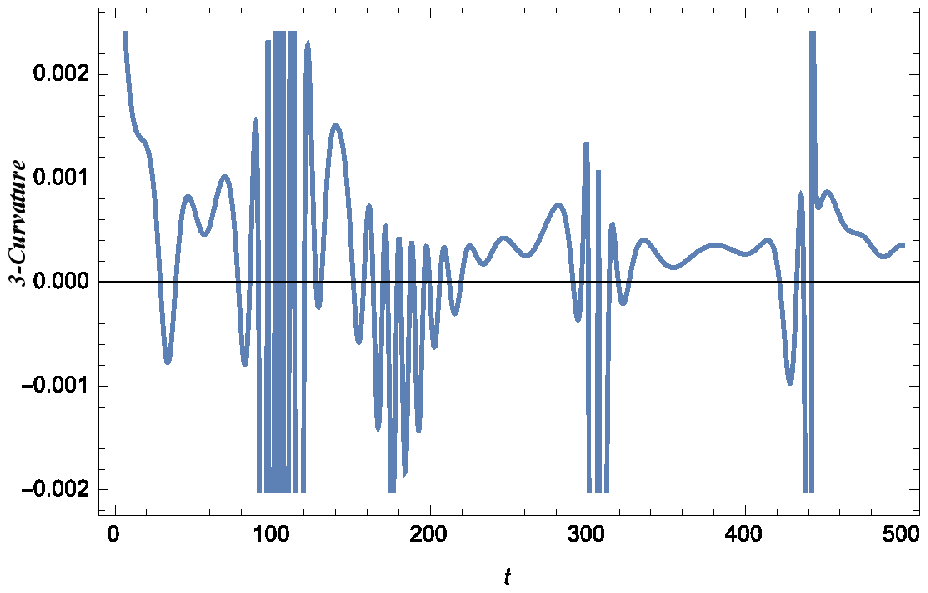}
\subcaption{\label{fig:shucurvature}}
\end{minipage}
\end{figure}

\begin{figure}[tbp]
\caption{Evolution of the shear (a) and curvature (b) in the type IX universe with negative $\Lambda $ and ghost field, with time, with initial conditions
differing by $\pm 0.001$, to illustrate the chaotic sensitivity of the
dynamics to small changes in initial data over many cycles of time evolution.
}\centering
\hfill \break 
\begin{minipage}{0.48\textwidth}
\includegraphics[width=1.0\linewidth, height=0.3\textheight]{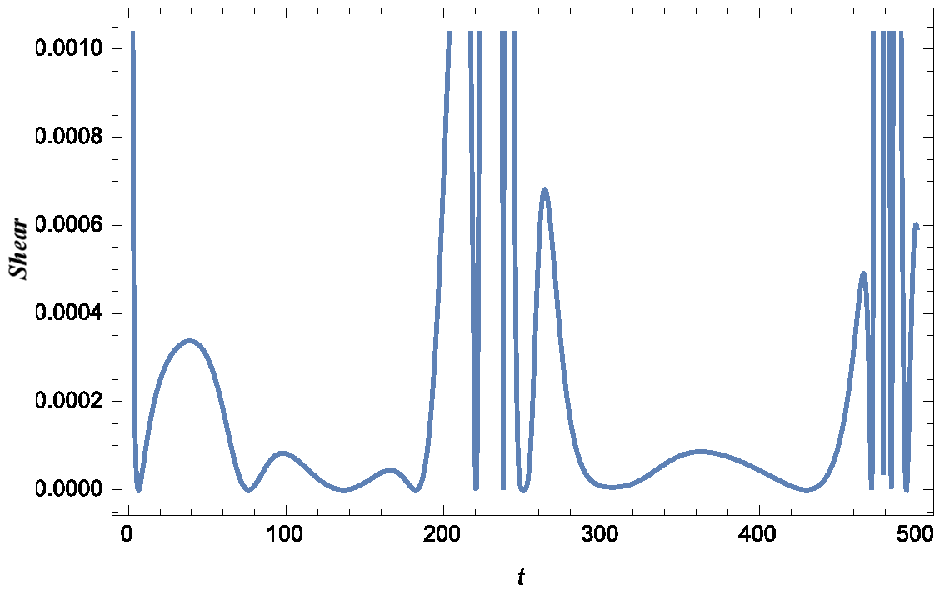}
\subcaption{\label{fig:shushear_ic_2}}
\end{minipage}\hfill 
\begin{minipage}{0.48\textwidth}
\includegraphics[width=1.0\linewidth, height=0.3\textheight]{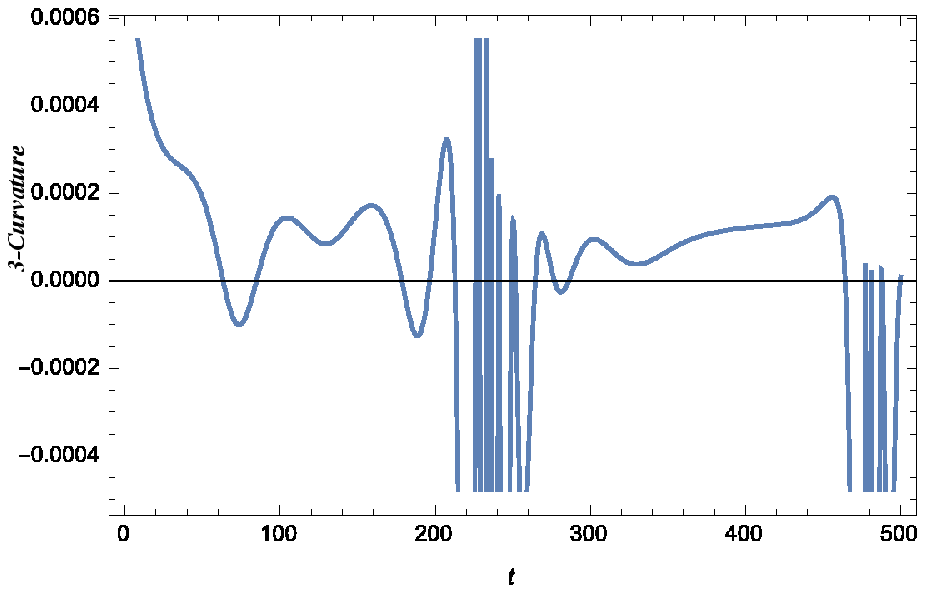}
\subcaption{\label{fig:shucurvature_ic_2}}
\end{minipage}
\end{figure}

Looking at Figures \ref{fig:shushear} and \ref{fig:shucurvature}, the shear
and the curvature also show several rapid and irregular oscillations as they
approach minima. On collapse of the model, they show no sign of approaching
isotropy, blowing up instead. To demonstrate that the long-term evolution is
chaotic, we find that on slightly changing the initial conditions by $\pm
0.001$ (where the initial conditions we have chosen for $[x,y,H]$ as well as
the densities of the ghost and radiation fields are of order 1), we find
that the number of oscillations in the shear and the curvature, as well as
their amplitude and shape, drastically change (especially for the initial
few cycles of the model). The behaviour for the shear for the second set of
initial conditions are shown in Figure \ref{fig:shushear_ic_2}.

Looking at the $3$-curvature next we find a similar situation, where the
shape of the oscillations as well as their frequency and amplitude vary
greatly with the same slight change in the initial conditions. We see this
in the Figure, \ref{fig:shucurvature_ic_2}.

It is interesting to note that this chaotic behaviour from cycle to cycle
occurs in $t$ time, rather than in its logarithms as is the case in the
chaotic behaviour seen on approach to the singularity if bounces do not
occur. \cite{bkl, jbchaos}.

\section{Conclusions}

We have investigated the fate of cyclic universes in the most general
spatially homogeneous closed universes of Bianchi type IX. These models are
considerably more general than those previously used to study oscillating
universes. We include radiation, matter and a stiff `ghost' field with
negative density to produce a smooth, non-singular bounce at finite volume
at the end of each cycle. A bounce at any finite volume, no matter how
small, avoids the issue of chaotic oscillations \cite{bkl} of the scale
factors on approach to the expansion minima. We investigate the effects of
an increase in radiation entropy from cycle to cycle, in accord with the
Second Law of thermodynamics, and also the effects of adding a positive or
negative cosmological constant to the Einstein equations. We find the
following:

\paragraph{Flatness and shape evolution:}

When $\Lambda =0$ we studied the evolution of type IX universes by analytic
approximations and by numerical evolution (starting with Kasner initial
conditions). The evolution follows an approximately axisymmetric form in
which different scale factors attain their maxima at different times before
turning around and collapsing towards their next minima. We found that as
the size of the volume maximum increases, the model approaches `flatness' in
the same way that isotropic closed universes do. However, the dynamics do
not approach isotropy as the volume maximum is approached or as the ensuing
minimum is approached. The successive expansion maxima grow increasingly out
of phase. The long-term dynamics are therefore anisotropic and differ
significantly from those predicted for inflationary universes.

\paragraph{General relativistic effects:}

The late-time evolution of type IX universe is dominated by intrinsically
general-relativistic effects associated with its 3-curvature anisotropy (for
which there is no Newtonian analogue). The sign of the 3-curvature scalar, $%
^{3}R$, can change with time. When the type IX universe is significantly
anisotropic, $^{3}R$ is negative and the dynamics cannot have an expansion
maximum. The universe therefore keeps on expanding and eventually becomes
close enough to isotropy for $^{3}R$ to become positive and then it is able
to experience a volume maximum and recollapse.

\paragraph{The effects of entropy increase:}

We injected radiation entropy at each finite expansion minimum to model the
effect of increasing entropy. We found that the entropy increase leads to
steady increase in the size of the volume maxima of successive cycles and to
their temporal duration but these maxima are anisotropic.

\paragraph{The effect of $\Lambda >0$:}

The addition of a cosmological constant is always found to bring the
oscillations in the volume of the universe to an end. This occurs no matter
how small the value of $\Lambda$ is. Oscillations of the universe occur, and
grow anisotropically as in the case of $\Lambda =0 $ until the size of the
maximum grows large enough for $\Lambda $ to become dynamically important
there. Subsequently, after a few scale factor transitory changes it will
dominate before any expansion maximum can occur and accelerate the expansion
towards an increasingly de Sitter-like metric evolution. This behaviour is
in accord with cosmic no-hair theorems even though, technically, they do not
apply to the type IX metric because it permits positive three-curvature,
which is excluded by the theorems

\paragraph{The effect of $\Lambda <0$:}

The addition of a negative cosmological constant causes any cosmological
model to recollapse, regardless of the sign of the 3-curvature. We use the
addition of $\Lambda <0$ to produce a simple bouncing model that experiences
a finite non-singular minimum at the end of each cycle because of the
presence of a ghost field. We follow the chaotic evolution of the scale
factors, the shear, and the 3-curvature from cycle to cycle. We showed that
there is sensitive dependence on initial conditions.

Our analysis introduced some simplifying assumptions. We consider only the
diagonal Bianchi IX metric with fluids that possess comoving velocity fields
and isotropic pressures. In a separate study, we will relax these
assumptions and show that a similar analysis is possible which leads to
similar conclusions. Thus we have shown that in the most general spatially
homogeneous anisotropic cyclic universes in general relativity with $\Lambda
=0$ the growth of entropy leads to never-ending cycles of increasing size
and duration, as Tolman first showed for isotropic models. However, although
these cycles approach flatness they do not approach isotropy and do not
resemble our observed universe. If we add $\Lambda >0$ then, no matter how
small the magnitude of $\Lambda $, the growing oscillations always come to
an end and subsequently the dynamics pass through a quasi-isotropic phase
before asymptoting towards the isotropic dynamics of a de Sitter metric.
These analyses can also be readily extended to other cyclic universe
scenarios that are based upon extensions of general relativity \cite{ekpy,
brand, lqg, q, BMK, bs, top, cap}.

\textbf{Acknowledgements} JDB is supported by the Science and Technology
Facilities Council (STFC) of the United Kingdom. CG is supported by the
Jawaharlal Nehru Memorial Trust Cambridge International Scholarship.

\end{document}